\documentclass[smallcondesed]{svjour3}
\usepackage{tikz}

\usepackage{comment}
\usepackage{cite}
\usepackage[shortlabels]{enumitem} 
\usepackage{amsmath}
\usepackage{url}
\usepackage{balance}
\newcommand{\bi}{\begin{itemize}[leftmargin=0.4cm]}
\newcommand{\ei}{\end{itemize}}
\newcommand{\be}{\begin{enumerate}}
\newcommand{\ee}{\end{enumerate}}
\newcommand{\tion}[1]{\S\ref{sect:#1}}
\newcommand{\fig}[1]{Figure~\ref{fig:#1}}
\newcommand{\eq}[1]{Equation~\ref{eq:#1}}
\setlist{nolistsep,leftmargin=5mm}

\usepackage{picture}
\usepackage{color}
\usepackage{colortbl}
\usepackage{xcolor}
\usepackage{listings}
\usepackage{commath}
\newenvironment{BLUE}{\color{black}}{\ignorespacesafterend}

\definecolor{lightgray}{gray}{0.8}
\definecolor{darkgray}{gray}{0.6}
\usepackage{tabu}

\definecolor{Gray}{gray}{0.95}
\definecolor{LightGray}{gray}{0.975}

\usepackage{marginnote}
\definecolor{darkgreen}{rgb}{0,0.3,0}

\usepackage{rotating}
\definecolor{Gray}{rgb}{0.88,1,1}
\definecolor{Gray}{gray}{0.85}
\definecolor{Blue}{RGB}{0,29,193}

\newcommand{\quart}[4]{\begin{picture}(100,4)%1
{\color{black}\put(#3,2){\circle*{4}}\put(#1,2){\line(1,0){#2}}}\end{picture}}
\newcommand{\WHERE}[1]{}%page \pageref{err:#1} (see \textcolor{blue}{#1})}

\newcommand{\HERE}[1]{}
\usepackage{wrapfig}

\definecolor{MyDarkBlue}{rgb}{0,0.08,0.45} 
\newenvironment{changed}{\par}{\par}

\newcommand{\ADD}[1]{#1}

\usepackage{times}

\def\baselinestretch{1}

\setlist{nosep}

 \usepackage[font={small}]{caption, subfig}

 \newcommand{\subparagraph}{} % defined before loading titlesec
\usepackage[compact,small]{titlesec}
\DeclareMathSizes{7}{7}{7}{7} 

\begin{document}  

\date{}
%
% --- Author Metadata here ---
% \conferenceinfo{FSE}{'15 Bergamo, Italy}
%\CopyrightYear{2007} % Allows default copyright year (20XX) to be over-ridden - IF NEED BE.
%\crdata{0-12345-67-8/90/01}  % Allows default copyright data (0-89791-88-6/97/05) to be over-ridden - IF NEED BE.
% --- End of Author Metadata ---

\title{Negative Results for Software Effort Estimation}


\author{Tim Menzies, Ye Yang, George Mathew,  Barry Boehm, Jairus Hihn
}

\institute{T. Menzies, G. Mathew \at
                CS, North Carolina State Univ., USA
                \email{tim.menzies@gmail.com}, \email{george.meg91@gmail.com}
            \and
            Y. Yang \at
                SSE, Stevens Inst., USA
                \email{ye.yang@stevens.edu}
            \and
            B. Boehm \at
                CS, Univ. of Southern California, USA
                \email{barryboehm@gmail.com}
            \and
            J. Hihn \at
                Jet Propulsion Laboratory/California Institute of Technology
                , USA
                \email{jairus.hihn@jpl.nasa.gov}
}
% There's nothing stopping you putting the seventh, eighth, etc.
% author on the opening page (as the 'third row') but we ask,
% for aesthetic reasons that you place these 'additional authors'
% in the \additional authors block, viz.
%\additionalauthors{Additional authors: John Smith (The Th{\o}rv{\"a}ld Group,
%email: {\texttt{jsmith@affiliation.org}}) and Julius P.~Kumquat
%(The Kumquat Consortium, email: {\texttt{jpkumquat@consortium.net}}).}
% Just remember to make sure that the TOTAL number of authors
% is the number that will appear on the first page PLUS the
% number that will appear in the \additionalauthors section.
\maketitle
\vspace{-1in}
\begin{abstract}
%\boldmath

 {\em Context:} More than half the literature on software effort estimation (SEE) focuses on 
comparisons of new estimation methods. Surprisingly, there are no studies comparing 
state of the art latest methods with decades-old approaches.

{\em Objective:} To check if new   SEE methods generated
better estimates than older methods.

{\em Method:} {\em Firstly}, collect effort estimation methods ranging from ``classical'' COCOMO
(parametric estimation over a pre-determined set of attributes) to ``modern'' (reasoning via analogy
using spectral-based clustering plus instance and feature selection, and \textcolor{black}{a recent ``baseline method'' proposed
in ACM Transactions on Software Engineering}).  {\em Secondly}, catalog the list of objections that lead to the development of
post-COCOMO estimation methods. {\em Thirdly}, characterize each of those objections as a
comparison between newer and older estimation methods. {\em Fourthly},  using four COCOMO-style
data sets (from 1991, 2000, 2005, 2010) and run those comparisons experiments.
{\em Fifthly}, compare the performance of the different estimators using a Scott-Knott procedure  using (i) the A12 effect size to
rule out ``small'' differences and (ii) a 99\% confident bootstrap procedure to
check for statistically different groupings of treatments). 

{\em Results:} The {\bf major negative results} of this paper are that for the COCOMO data sets, nothing we studied did any 
better than Boehm's original  procedure.  

{\em Conclusions:} When COCOMO-style attributes are available, we strongly recommend (i)~using
that data and (ii)~use COCOMO to generate predictions.
We say this since the experiments
of this paper show that, at least for effort estimation,
{\em how data is collected} is more important than {\em what learner is applied to that data}.

\end{abstract}

\vspace{1mm}
\noindent
{\bf Categories/Subject Descriptors:} 
D.2.9 [Software Engineering]: Time Estimation;
K.6.3 [Software Management]: Software Process

\vspace{1mm}
 
\newpage
\section{Introduction}
This paper is about a negative result in software effort estimation-- specifically:
\bi
\item
For project data
expressed in a certain
way (the COCOMO format~\cite{boehm00b});
\item
Despite decades of work into 
alternate methods;
\item
Best predictions  from that data
come from a parametric  method proposed in 2000~\cite{boehm00b}.\ei
This conclusion comes with two
caveats. 
  Firstly, not all projects can be expressed in terms of COCOMO--
  but when there is a choice, the results
  of this paper argue that there is
  value in using that format.
Secondly, our conclusion is about {\em solo} prediction methods which is different to the  
{\em ensemble} approach~\cite{Minku2011,Minku2013,koc11a,Minku16b}-- but if using ensembles, this paper shows that parametric estimation would be   a viable  
 ensemble member.

For pragmatic and methodological reasons,it is important to report   negative results like the one described above. Pragmatically, it is important
for industrial practitioners to know that (sometimes) they do not need to waste time straining to understand  bleeding-edge technical papers. In the following, we precisely define the class of project data that {\em does not} respond well to  bleeding-edge 
effort estimation techniques.
For those kinds of data sets,   practitioners can be rest assured that it is reasonable and responsible and useful to use simple
traditional methods.

Also, methodologically, it is important to acknowledge mistakes. According to 
 Karl Popper (a prominent figure in the philosophy of science~\cite{popper63}), the ``best'' theories are  the ones that have best survived vigorous debate.  Having been engaged in some
 high-profile debates (in the field
of software analytics~\cite{me07e}), we assert that such criticisms are very  useful
since they help a researcher (1)~find flaws in old ideas and (2)~evolve better
new ideas. That is, finding and acknowledging mistakes should be regarded as a routine part of
standard operations procedure for science.

\begin{BLUE}\HERE{Reviewer1a} 
Given the above, it is troubling  that 
there are very few negative results   in the field of software analytics. 
What does happen, occasionally,  are  reports of small
corrections to prior work. Given the complexity of software analytics, this absence of such failure reports is highly suspicious. For examples of such reports, see (e.g. as done in \cite{menzies13err,Murphy-Hill2012}).
\end{BLUE}

Why are these reports so rare? There are many possible reasons and here we speculate on two possibilities.
Firstly, such negative reports may not be acknowledged as ``worthwhile'' by the community. Forums such
as this special issue are very rare (which is why this issue is so important).
Secondly, it is not standard practice in software analytics for researchers to benchmark their latest
results against some supposedly simpler ``straw man'' method. In his textbook on {\em Empirical Methods in AI}, Cohen~\cite{cohen95} strongly advises such ``straw man'' comparisons, since sometimes, they reveal that the supposedly
superior method is actually overly complex. Hence it
always useful  to compare methods
against simpler alternative. 

\begin{BLUE}\HERE{Reviewer1b} That said,
in some  cases no such method is available making such  benchmarking impossible. Although as a   domain starts to become more mature,
these comparisons can be conducted; see, e.g. the
many experiments on defect prediction~\cite{Scanniello2013} or tag inference for Stack Overflow posts~\cite{stanley2013predicting}.
\end{BLUE}
Accordingly, this paper checks an interesting,
but currently unexplored aspect of effort estimation.
We check
if there exists data sets from which  very old methods do just as well as anything else. 
We consider data expressed in terms of the COCOMO ontology: 23 attributes describing
a software project,  as well as aspects of its personnel, platform and product features\footnote{For full details on these attributes, see {\S}4 of this paper.}. We will show
that (given this diverse sample of data types collected from a project)   Boehm's 2000 model works
as well (or better) than everything else we tried.  Hence, we strongly recommend that if that kind
of data is available, then it should be collected and it should be processed using Boehm's 2000 COCOMO model.

To guide our exploration,
 this paper asks four research questions.  These
 questions have been selected based  on our experience debating
 the merits of COCOMO vs alternate methods. Based on our experience,
 we assert that each of the following questions has been used to motivate
 the development of some alternate to the standard COCOMO-II model:

~{\bf RQ1: Is parametric estimation no better than
   using just Lines of Code measures?} 
  (an   often heard, but rarely tested, comment).

 \begin{figure}[!b]
\begin{minipage}{2.3in}\noindent{\scriptsize
\begin{tabular}{r|@{~}r|@{~}r|@{~}r|@{~}l}

Types of projects&\begin{sideways}COC81\end{sideways} & \begin{sideways}NASA93\end{sideways}& \begin{sideways}COC05\end{sideways} &\begin{sideways}NASA10\end{sideways}\\\hline 
Avionics&     &26&10&17\\\hline
Banking&       &      &13&     \\\hline
Buss.apps/databases&7&4&31&  \\\hline   
Control&9&18&13&     \\\hline
Human-machine interface&12&       &       &     \\\hline
Military, ground&       &       &8&     \\\hline
Misc&5&4&5&     \\\hline
Mission Planning&      &16&       &     \\\hline
SCI scientific application&16&21&11&     \\\hline
Support tools, &7&        &       &   \\\hline  
Systems&7&3&2&

%% NASA10& COC05 & NASA93& COC81 & Types of projects\\\hline\hline
%%      &     5 &4      &       & Misc\\\hline
%%      &       &        &  7   &  Support (tools, utilities, etc)\\\hline
%%      &      8&       &       & Military, ground\\\hline
%%      &      13&      &       & Banking\\\hline
%%      &      2&3      & 7     & Sys (OS, compilers, sensors,etc)\\\hline
%%      &      31&4     &  7    & Business apps/data processing\\\hline
%%      &       &16      &      & Mission Planning\\\hline
%%      &     13  &18     &   9 & Control\\\hline
%%      &       &       & 12    & Human=machine interaction\\\hline
%%      &      11& 21     & 16   & SCI scientific application\\\hline
%%  17  &     10 &26      &     & Avionics Monitoring
\end{tabular}}
\end{minipage}~\begin{minipage}{2.3in}

\noindent\includegraphics[width=2.2in]{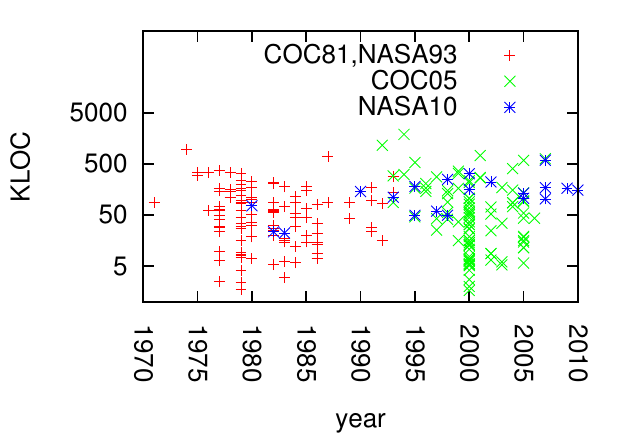}
\end{minipage}
\caption{Projects used by the learners in this study. \fig{cparems}
shows project attributes. 
COC81 is the original data from 1981 COCOMO book~\cite{boehm81}. 
This comes from projects dating 1970 to 1980.
NASA93 is NASA data collected  in the early 1990s
 about software that supported  the planning activities for the International
Space Station. 
Our other data sets are  NASA10 and COC05 (the latter is
proprietary and 
cannot be released to the research community). 
The non-proprietary data  (COC81 and NASA93 and NASA10) is available at
http://openscience.us/repo or in \fig{nasa10}.
}\label{fig:types}
\end{figure}

\begin{figure}[!b]
\begin{center}
\includegraphics[width=3.6in]{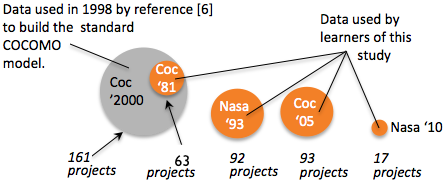}
\end{center}
\caption{\ADD{Projects in this study. COC81 is a subset of COCOMO-II.  Note that NASA'93 and
COC'05 and NASA'10 have no
overlap with the data used to define the version of COCOMO
used in this paper. }}\label{fig:dataused}
\end{figure} 

  \begin{figure}[!t]
  \begin{BLUE}
  \scriptsize
  \begin{tabular}{|@{}r@{~}r@{~}r@{~}r@{~}r@{~}|@{}r@{~}r@{~}r@{~}r@{~}r@{~}r@{~}r@{~}r@{~}r@{~}r@{~}r@{~}r@{~}r@{~}r@{~}r@{~}r@{~}r@{~}|r@{~}|r|}\hline
   
prec&flex&resl&team&pmat&rely&cplx&data&ruse&     time&stor&pvol&acap&pcap&pcon&aexp&plex&     ltex&tool&sced&site&docu&kloc&months\\\hline
2&2&2&3&3&4&5&4&3&5&6&4&4&4&3&4&3&3&1&3&4&4&77&1830\\
2&2&2&3&3&5&5&2&3&5&6&2&4&3&3&2&1&2&2&3&4&4&24&648\\
2&2&2&3&3&4&5&3&3&5&5&4&3&3&3&3&2&2&1&3&4&4&23&492\\
2&2&3&3&2&4&4&3&2&3&3&4&3&3&3&3&3&4&2&3&5&3&146&3292\\
2&3&3&5&3&3&4&3&2&4&4&2&5&5&4&5&1&5&3&3&6&3&113&1080\\
3&3&3&3&3&3&4&3&2&3&3&3&3&3&3&4&3&4&2&3&4&3&184&1043\\
5&3&3&3&4&4&4&3&2&3&3&2&3&3&3&5&3&4&2&3&5&3&61&336\\
5&3&3&4&4&4&5&3&2&3&3&2&3&3&3&5&3&4&2&3&6&3&50&637\\
3&3&3&2&3&4&5&3&2&3&3&3&3&3&3&4&3&4&2&3&5&3&253&2519\\
3&3&4&3&3&4&4&3&4&3&3&2&3&4&3&3&1&4&5&3&2&3&159&1048\\
3&3&3&3&3&4&5&3&2&3&3&4&4&4&5&4&4&4&2&1&5&3&324&1735\\
3&2&4&4&3&4&5&3&4&3&4&5&4&4&3&4&4&3&4&2&6&3&224&691\\
5&2&2&4&3&4&3&3&4&5&4&3&4&4&3&4&4&4&3&3&3&3&105&320\\
3&2&2&4&3&4&3&3&3&3&3&2&4&4&3&4&4&4&3&3&3&3&173&329\\
3&2&4&3&3&4&5&3&4&3&3&4&3&4&4&4&3&3&3&3&5&3&597&1705\\
4&2&4&3&5&4&3&2&3&3&4&4&2&2&3&3&5&5&3&3&5&3&155&789\\
4&3&3&3&4&4&4&3&2&3&3&3&4&4&3&5&4&4&2&3&5&3&170&552\\\hline 
 \multicolumn{5}{|c|}{scale factors}&\multicolumn{17}{c|}{effort multipliers}&\multicolumn{1}{r|}{size}&\multicolumn{1}{r|}{effort}\\\hline 
  \end{tabular}
  \HERE{Reviewer2b}
  \caption{\textcolor{black}{ The 17 projects in NASA10 (one row per project). For a definition of the terms in row1 (``prec'', ``flex'', ``resl'' etc.) see \fig{cparems}.
    As to the different columns, scale factors change effort exponentially while effort multipliers have a linear impact on effort.
    Any effort multiplier with a value of ``3'' is a {\em nominal} value; i.e. it multiplies the effort by a multiple of 1.0. Effort multipliers
    above and below ``3'' can each effect project effort by a multiple ranging from 0.7 to 1.74.  For full details on how these values are used,
    see \fig{coc2}.}}\label{fig:nasa10}
  \end{BLUE}
\end{figure}

 \begin{figure}[!t]
{\scriptsize
\begin{center}
\begin{tabular}{|p{0.2in}|p{1.46in}|p{0.77in}|p{0.77in}|p{0.77in}|}\hline

 & Definition & Low-end = \{1,2\}
 &Medium =\{3,4\} &High-end= \{5,6\} \\\hline

\multicolumn{1}{c}{~}\\

\multicolumn{5}{l}{Scale factors:}\\\hline
Flex   &  development flexibility   & development process
rigorously defined & some guidelines, which can be relaxed & only
general goals defined\\\hline

Pmat    & process maturity  &  CMM level 1 &   CMM level 3  &  CMM level 5 \\\hline

Prec & precedentedness  &  we have never built this kind
of software before &    somewhat new &
thoroughly familiar \\\hline

Resl &  architecture or risk resolution  &  few interfaces
defined or few risks eliminated  &  most interfaces defined or most
risks eliminated   & all interfaces defined or all risks
eliminated\\\hline

Team  &   team cohesion  &  very difficult interactions &
basically co-operative  &  seamless interactions\\\hline

\multicolumn{1}{c}{~}\\

\multicolumn{5}{l}{Effort multipliers}\\\hline
acap  &  analyst capability  &  worst 35\% &   35\% - 90\% &  best 10\% \\\hline

aexp   &  applications experience  &  2 months &   1 year  &  6 years\\\hline

cplx   &  product complexity   & e.g. simple read/write
statements & e.g. use of simple interface widgets  &  e.g.
performance-critical embedded systems\\\hline

data   &  database size 
(DB bytes/SLOC) &
10 & 100 &    1000 \\\hline

docu   &  documentation   & many life-cycle phases not
documented      & &  extensive reporting for each life-cycle phase\\\hline

ltex   &  language and tool-set experience   & 2 months  &  1
year & 6 years \\\hline

pcap   &  programmer capability  &  worst 15\%   & 55\%  &  best 10\% \\\hline

pcon   &  personnel continuity \newline
(\% turnover per year) &
    48\% &    12\%  & 3\% \\\hline

plex   &  platform experience  &  2 months  &  1 year  &  6 years\\\hline

pvol   &  platform volatility ($\frac{frequency~of~major~changes}{frequency~of~minor~changes}$) &
$\frac{12~months}{1~month}$   & $\frac{6~months}{2~weeks}$ &
$\frac{2~weeks}{2~days}$\\\hline

rely   &  required
reliability &   errors are slight inconvenience  &  errors are easily
recoverable   & errors can risk human life\\\hline

ruse   &  required
reuse &   none &    multiple program  & multiple product lines\\\hline

sced  &   dictated development\newline schedule &    deadlines moved to
75\% of the original estimate &  no change
&  deadlines moved back to  160\% of original estimate\\\hline

site   &  multi-site development   & some contact: phone, mail&
some email  &  interactive multi-media\\\hline

stor  &   required \% of available
RAM & N/A
 &   50\% &  95\% \\\hline

time  &   required \% of available CPU &
N/A&     50\%
   &  95\% \\\hline

tool   &  use of software tools  &  edit,code,debug &&
integrated with life cycle\\\hline

\multicolumn{1}{c}{~}\\

\multicolumn{5}{l}{Effort}\\\hline

months & construction effort  in months& \multicolumn{3}{l|}{1 month =  152 hours (includes development \& management
hours).  
}\\\hline
\end{tabular}
\end{center}
} \caption{COCOMO-II attributes.}
\label{fig:cparems}
\end{figure}

~{\bf RQ2: Has parametric estimation been superseded
 by more recent estimation methods?}  We  
 apply our ``best''
learner, as well as 
case-based reasoning and regression trees.

~{\bf RQ3: Are the old parametric tunings irrelevant
  to more recent projects?}  
  COCOMO models are learned by ``tuning''
  the default model parameters using local
  project data. COCOMO-II shipped with a set
  of parameters learned from a particular set of
  projects from 1995 to 2000. We apply those 
COCOMO-II tunings, without modification, to a wide range of projects dating from
1970 to 2010.

~{\bf RQ4: Is parametric estimation expensive to deploy  at some new site?}
We try   tuning estimation models on  small training sets 
as well as simplifying the specification of  projects.

%~{\bf RQ5: Are parametric estimates unduly sensitive to errors in the size estimate?%}
%In the context of RQ4, we check what happens
%if there are large errors in the ``thousand lines of code''(KLOC) estimate.

To explore these questions, we use  COCOMO since its
internal details have been fully
published~\cite{boehm00b}. Also, we can access a full implementation of the  2000
COCOMO model.
Further, we have access to numerous interesting  COCOMO data
sets: see \fig{types} and \fig{dataused}.
\begin{BLUE}
\HERE{Reviewer2a}
With one exception, our learning experiments do not use the data
that generated   standard COCOMO.
That exception is the  COC81 data-- which  lets us  compare new methods
against the  labor intensive methods used to make standard COCOMO-- see 
\fig{dataused}.
\end{BLUE}

Using that data,  the experiments of this paper conclude that
the answer to all  our  research questions is nearly always
``no''.  
The RQ1 experiments show that good estimates use many variables
and  poorer estimates result from   some trite calculation based on KLOC.   
%Hence, as seen in RQ5,  we can make some degree of
%error in our KLOC estimates without damaging the overall estimation process (the cav%eat here is that if the KLOC
%estimates grow too erroneous-- more than 200\%--
%then some degradation is seen).%
%
As to the other research questions (RQ2, RQ3, RQ4), those results imply that 
the continued
use of parametric estimation can still be endorsed-- at least for data expressed
in terms of the 23 COCOMO attributes. 

For a sample of our data  see the NASA10 data set in \fig{nasa10}.

 \section{About Effort Estimation}
 \subsection{History}

Accurately estimating software development
effort  is of vital
importance. 
Under-estimation can cause schedule and budget
overruns as well as project
cancellation~\cite{CLCS03}.  Over-estimation delays
funding to other promising ideas and
organizational competitiveness~\cite{koc11a}.
%XXX history of parametric models learned via regression
Hence, there is a long history
of researchers exploring software effort estimation; e.g. \cite{wol74,frei79,putnam76,black77,herd77,watson77,jensen83,park88,boehm81,Walkerden1999,shepperd97,jorgensen05,me06d,burgess01}.
In 2007, Jorgensen and Shepperd
reported on hundreds of research papers dating back to the 1970s devoted to
the topic, over half of which propose some innovation
for developing new estimation
models~\cite{jorgensen05}. Since then,
many such papers have been published;
e.g. \cite{lokan06,cora10,minku14,Li2007,Li2009a,keung2008a,keung2008b,keung2008c,koc11b,me12a,me13a,kocaguneli2014transfer}.

In the 1970s and 1980s, this kind of research was focused on
{\em parametric estimation} as done 
by Putnam and
others~\cite{wol74,frei79,black77,herd77,watson77,boehm81}. For example, Boehm's
COnstructive COst MOdel (COCOMO)
model~\cite{boehm81} 
 assumes  that effort varies exponentially on size as seen in this parametric form:
$\mathit{effort} \propto \mathit{a \times KLOC}^b$. To deploy this equation in an organization,
local project data is used to tune the  $(a,b)$ parameter values, If local
data is unavailable, new projects can reuse prior tunings,  with  minor
tweaks~\cite{me04h}. 
COCOMO is a parametric method; i.e. it is a 
{\em model-based} method that (a)~assumes that the target model has a particular structure,
then (b)~uses model-based methods to fill in the details of that structure (e.g. to set some tuning parameters).

%XXX what if not in cocomo format. answer : then use something else. caution:
%XXX the results thos this paper suggest that if possible& better estimates might be obtained from
%XXX starting with cocomo- but this is uaully a domain descision (sometimes local experts
%just prefer using their local in-house descriptors.

Since that work on parametric estimation, researchers
have innovated other methods based on
regression
trees~\cite{shepperd97}
case-based-reasoning~\cite{shepperd97}, spectral
clustering~\cite{me12d}, genetic
algorithms~\cite{cordero97,burgess01}, etc.  These methods
can be augmented with  ``meta-level'' techniques like tabu search~\cite{cora10}, feature selection~\cite{chen05}, instance selection~\cite{koc11b},
feature synthesis~\cite{me12a}, active learning~\cite{me13a}, transfer learning~\cite{kocaguneli2014transfer}.
temporal learning~\cite{lokan09,minku14}, and many more besides.

%% This paper comments on a curious disconnect between the academic research and the commercial effort estimation industry.

%% in the software effort
%% estimation is one of the oldest, most enduring,
%% research themes in SE
%% Effort estimation is still the focus
%% of much research activity.  
%% Since the 1970s, one of us (Boehm) has lead an large consortium
%% of researchers building and revising the
%% COnstructive COst
%% MOdel (COCOMO) model. First published in 1981~\cite{boehm81},
%% the model was updated to COCOMO-II~\cite{boehm00b}. A newer COCOMO-III model is also
%% currently
%% under development~\cite{rosa14}. 
 
\subsection{Current Practice}
In her keynote address to ICSE'01, Mary Shaw~\cite{shaw01} noted that it can take up to a
decade  for  research innovations
to become stable and then another decade after that to become widely popular. Given that, it would be reasonable
to expect commercial adoption of  the 1990s estimation work
on  regression trees~\cite{shepperd97} or case-based-reasoning~\cite{shepperd97}. However, 
this has not happened.
Parametric estimation is
widely-used, especially across the aerospace
industry and various U.S. government agencies. For example:
\bi
\item
NASA routinely checks  software estimates 
in  COCOMO~\cite{dabney07}.  
\item
In our work with the Chinese and the United States software industry,
we saw an   almost exclusive
use  of parametric estimation tools such as those offered by 
Price Systems (pricesystems.com) and  Galorath (galorath.com).
\item
Professional societies, handbooks and
certification programs are mostly developed around 
parametric estimation methods and tools; e.g. see the 
International Cost Estimation and Analysis Society; the
NASA Cost Symposium;  the
International Forum on COCOMO and Systems/Software
Cost Modeling (see the websites \url{http://tiny.cc/iceaa}, \url{http://tiny.cc/nasa_cost}, \url{http://tiny.cc/csse}).
\ei

%% These new model  are of many types.
%% Older approaches
%% used a fixed set of parameters (e.g
%%  COCOMO, FPA~\cite{albrecht79},
%% SLIM~\cite{putnam80}).
%% More recent work has applied methods
%% that can accept a wider range  parameters including
%% analogy-based methods~\cite{shepperd97} and 
%% combination methods that take advantage of
%% multiple simpler methods. For example, Corazza
%% et al.~\cite{cora10} uses tabu search to configure support vector
%% machines for effort estimation.
%% Other approaches  combine
%% hundreds of different methods-- see
%%  Menzies et al.~\cite{me06d} or 
%% Kocaguneli~\cite{me11a}.
%% Yet other approaches use
%% temporal learning methods  incrementally modify past
%% models or data to make estimates on new projects (see the 
%% pioneering work of Lokan and Mendes~\cite{lokan06}, or more recent
%% work by Minku and Yao~\cite{minku14})).
%% XXX implications for more researh
%% {\scriptsize
%% \Tree [.Projects\ can\\have\ COCOMO\\attributes? 
%%            [.yes [.Dozens\ of\\examples? 
%%                      [.yes COCOMO ]
%%                      [.no  COCOMO-II\\+\ COCOUUT\\+\ Column\ prune ] ] ] 
%%            [.no  [.Dozens\ of\\examples? 
%%                       [.yes Try\ parametric\\methods\ first ]
%%                       [.no  Case-based\\reasoning ] ] ] ]
%% }

\subsection{But Does Anyone Use COCOMO?}

Two of the myths of effort estimation is that 
(1)~no one uses model-based estimation like COCOMO;
and (2)~estimates are always better done using expert-based guess-timation.

These myths are misleading.
As seen above,  model-based
parametric methods are  widely used in industry and
are strongly advocated by professional societies.
Also, 
while it is true that expert-based estimation is a common practice~\cite{boehm00a}, this is not to say that this should be recommended as the {\em best} or {\em only} way to make estimates:
\bi 
\item
Jorgensen~\cite{Jorgensen2004} reviews studies 
comparing  model- and expert- based estimation and concludes that there
there is no clear case that expert-methods are better.
\item
In 2015, Jorgensen further argued~\cite{jorg15} that   model-based methods are useful for learning the  {\em uncertainty}
about particular estimates; e.g.
by running those models many times, 
each time applying
small mutations to the input data.
\item 
Valerdi~\cite{valerdi11} lists the
cognitive biases that can make an expert offer poor expert-estimates.
\item Passos et al. show that many
commercial software engineers generalize from their
first few projects for all future
projects~\cite{passos11}.
\item
Jorgensen \& Gruschke~\cite{jorgensen09} document how
  commercial  ``gurus'' rarely use lessons
  from past projects to improve their future expert-estimates. 
 They offer examples where this
  failure to revise prior beliefs   leads to poor
 expert-based estimates.
  \ei 
Much research has concluded that the best estimations come from {\em combining} the predictions
from {\em multiple oracles}~\cite{koc11a,chulani99,baker07,valerdi11}.  
Note that it is much easier to apply this double-check strategy using expert+model-based methods 
than by comparing the estimates from multiple expert teams.
For example, all the model-based methods  studied in this paper can generate estimates
in just a few seconds. In comparison, expert-based estimation is orders of magnitude slower-- as seen in  
Valerdi's  
COSYSMO expert-method.
While a strong proponent of this approach, Valerdi concedes that 
``(it is)  extremely time
consuming when large sample sizes are needed''~\cite{valerdi11}.
For example, he once
recruited 40 experts to three expert sessions, each of which ran for three hours.
Assuming a 7.5 hour day,
then that study took $3*3*40 /7.5 = 48\; \mathit{days}$.

COSYSMO is an elaborate expert-based method. An alternate, more lightweight expert-method is  ``planning poker''~\cite{molokk08} where
participants offer anonymous
``bids'' on the 
completion time for a project. If  the bids are widely divergent, then the factors
leading to that disagreement are elaborated and debated. This cycle of bid+discuss continues
until a consensus has been reached.

While planning poker is widely advocated in the agile community,
there are surprisingly few studies assessing this method (one rare exception is~\cite{molokk08}).
Also,   planning poker is used to assess effort
for particular tasks in the scrum backlog-- which is a different and simpler task
than the {\em initial} estimation of  large-scale
projects. This is an important issue since, for larger
projects, the initial budget allocation may require a significant amount of intra-organizational lobbying between groups with competing concerns. For such large-estimate-projects, it can
be challenging to change the initial budget allocation. Hence, it is important to get
the initial estimate as accurate as possible.

\subsection{COCOMO: Origins and Development}
These concerns with  expert-based estimation  date
back many decades and were the genesis for  COCOMO. In 1976, Robert Walquist (a TRW division general manager)
told  Boehm: \begin{quote}{\em ``Over the last three
weeks, I've had to sign proposals that committed us
to budgets of over \$50 million to develop the
software.  In each case, nobody had a good
explanation for why the cost was \$50M vs. \$30M or
\$100M, but the estimates were the consensus of the
best available experts on the proposal team.  We
need to do better. Feel free to call on experts
\& projects with data on previous software cost.''}\end{quote}

TRW had a previous model that worked well for a part
of TRW's software business~\cite{wol74}, but it
did not relate well to the full range of embedded
software, command and control software, and
engineering and scientific software involved in
TRW's business base.  Having access to experts and
data was a rare opportunity, and a team involving
Ray Wolverton, Kurt Fischer, and Boehm conducted a
series of meetings and expert exercises to find
the relative significance of various  cost
drivers. Combining local expertise  and data, plus some prior results 
such as~\cite{putnam76,black77,herd77,watson77},  and early versions of the RCA
PRICE S model~\cite{frei79}, a model called SCEP was created (Software Cost
Estimation Program).
Except for
one explainable outlier, the estimates for
 20 projects with solid data were within 30\% of
the actuals, most within 15\% of the actuals.

\begin{figure}[!t]
\begin{center}
\begin{lstlisting}
_  = None;  Coc2tunings = [[
#              vlow  low   nom   high  vhigh  xhigh   
# scale factors:
'Flex',        5.07, 4.05, 3.04, 2.03, 1.01,    _],[
'Pmat',        7.80, 6.24, 4.68, 3.12, 1.56,    _],[
'Prec',        6.20, 4.96, 3.72, 2.48, 1.24,    _],[
'Resl',        7.07, 5.65, 4.24, 2.83, 1.41,    _],[
'Team',        5.48, 4.38, 3.29, 2.19, 1.01,    _],[
# effort multipliers:        
'acap',        1.42, 1.19, 1.00, 0.85, 0.71,    _],[
'aexp',        1.22, 1.10, 1.00, 0.88, 0.81,    _],[
'cplx',        0.73, 0.87, 1.00, 1.17, 1.34, 1.74],[
'data',           _, 0.90, 1.00, 1.14, 1.28,    _],[
'docu',        0.81, 0.91, 1.00, 1.11, 1.23,    _],[
'ltex',        1.20, 1.09, 1.00, 0.91, 0.84,    _],[
'pcap',        1.34, 1.15, 1.00, 0.88, 0.76,    _],[ 
'pcon',        1.29, 1.12, 1.00, 0.90, 0.81,    _],[
'plex',        1.19, 1.09, 1.00, 0.91, 0.85,    _],[ 
'pvol',           _, 0.87, 1.00, 1.15, 1.30,    _],[
'rely',        0.82, 0.92, 1.00, 1.10, 1.26,    _],[
'ruse',           _, 0.95, 1.00, 1.07, 1.15, 1.24],[
'sced',        1.43, 1.14, 1.00, 1.00, 1.00,    _],[ 
'site',        1.22, 1.09, 1.00, 0.93, 0.86, 0.80],[ 
'stor',           _,    _, 1.00, 1.05, 1.17, 1.46],[
'time',           _,    _, 1.00, 1.11, 1.29, 1.63],[
'tool',        1.17, 1.09, 1.00, 0.90, 0.78,    _]]

def COCOMO2(project,  a = 2.94, b = 0.91, # defaults
                      tunes= Coc2tunings):# defaults 
  sfs,ems,kloc   = 0, 5 ,22        
  scaleFactors, effortMultipliers = 5, 17
  
  for i in range(scaleFactors):
    sfs += tunes[i][project[i]]
    
  for i in range(effortMultipliers):
    j = i + scaleFactors
    ems *= tunes[j][project[j]] 
    
  return a * ems * project[kloc] ** (b + 0.01*sfs) 
\end{lstlisting}
\end{center}
\caption{COCOMO-II: effort estimates from a {\em project}.
Here, {\em project} has  5 scale
factors plus 17 effort multipliers plus KLOC. 
``Xhigh'' is show for ``extremely high''.
Each attribute except KLOC and effort is scored
using the scale very low = 1, low=2, up to
xhigh=6.
Note all attributes extend across the entire
range very low to extremely high since,
in Boehm's modeling work, not all effects extend
across the entire range.
For an explanation of the attributes shown in
green, see \fig{cparems}.}\label{fig:coc2}
\end{figure}

%%  and worked with projects to gather uniform
%% software cost driver and effort data on its 1970s
%% large software projects.  The data definitions were
%% based on TRW’s waterfall-model based software
%% development policies and standards. They also
%% benefited from a concurrent surge in publications
%% describing software cost estimation models such as~\cite{putnam76,black77,herd77,watson77} and early versions of the RCA
%% PRICE S model~\cite{frei79}.
%% SCEP  became standard practice for
%% large TRW projects; and a TRW Office of Software
%% Cost Estimation was established to provide its usage
%% support, training, data collection and analysis, and
%% evolution, which continues to this day.

%% The conclusion for that work
%% was that a reasonably accurate model could
%% be developed, but that proposals and projects should
%% complement its results with expert-judgment based
%% estimates, and reconcile their results where
%% necessary.  
%% It was given the name SCEP, for Software Cost
%% Estimation Program; its use along with expert
%% judgment estimates became standard practice for
%% large TRW projects; and a TRW Office of Software
%% Cost Estimation was established to provide its usage
%% support, training, data collection and analysis, and
%% evolution, which continues to this day.

%% SCEP was highly successful in proposal and
%% management use, but it was proprietary to TRW and
%% could not be used by customers to evaluate other
%% companies' cost estimates.  Thus, TRW became
%% receptive to having a version that generalized
%% beyond TRW experience but was still accurate for
%% TRW.  

After  gathering some further data from subsequent
TRW projects and about 35 projects from teaching
software engineering courses at UCLA and USC along
with commercial short courses on software cost
estimation, Boehm was able to gather 63 data points
that could be published and to extend the model to
include alternative development modes that covered
other types of software such as business data
processing.  The resulting model was called the
COnstructive COst MOdel, or COCOMO, and was
published along with the data in the book Software
Engineering Economics~\cite{boehm81}. 
In COCOMO-I, project attributes
were scored using just a few coarse-grained values (very low,
low, nominal, high, very high). These attributes
are {\em effort multipliers} where
a off-nominal value changes the estimate by some number
greater or smaller than one.
In COCOMO-I, all attributes (except KLOC)
influence effort in a linear manner.

Following the release of COCOMO-I Boehm created a consortium for
industrial organizations using COCOMO .
The consortium
collected information on 161 projects from commercial,
aerospace, government, and non-profit organizations.
Based on an analysis of those 161 projects, Boehm
 added  new attributes called {\em scale factors}
that had an {\em exponential impact}
on effort (e.g. one such attribute was process maturity).
Using that new data, Boehm and his colleagues developed
the  {\em tunings} shown in \fig{coc2} that
map the project descriptors (very low, low, etc)
into the specific values used in the COCOMO-II model
(released in 2000~\cite{boehm00b}):
\begin{equation}\label{eq:cocII}
\mathit{effort}=a\prod_i EM_i *\mathit{KLOC}^{b+0.01\sum_j SF_j}
\end{equation}

Here, {\em EM,SF} are  effort multipliers and scale
factors respectively and
 $a,b$ are the {\em local calibration} parameters (with default values of 2.94 and 0.91).
Also, {\em effort}
measures ``development months'' where one month
is 152 hours of work  (and includes development and management hours).
For example, if {\em effort}=100, then according to COCOMO,
five developers would finish
the project in 20 months.

%% To confirm this second possibility we ran two small experiments
%% with the COCOMO equation. In both experiments,
%% project size was varied from 0 to 1,000,000 lines of code.
%% For each size, two estimates were generated using 
%% by adjusting the effort/scale factors to their (1)~{\em least}
%% and (2)~{\em most} values. As  shown in \fig{linear},
%% the exponent on the KLOC term is very close to one while,
%% in the worst case (on the orange curve) other factors
%% can change the estimate by three orders of magnitude (as witnessed
%% by the leading 1699.07 value). 
%% \end{BLUE}

%% \begin{figure}[!h]
%% \includegraphics[width=5in]{linear.png}
%% \caption{\textcolor{blue}{COCOMO estimates increase
%% as very low-order power series.
%% Assuming {\em most} / {\em least} values for
%% effort and scale factors, the exponent
%% on the KLOC term is 0.9723 and 1.23, respectively.}}\label{fig:linear}
%% \end{figure}
 
\subsection{COCOMO and Local Calibration}\label{sect:coconut}
COCOMO models are learned by ``tuning''
  the default model parameters using local
  project data.
When local data is scarce, approximations can be used to
tune a model using just a handful of examples.  

For example,
 COCOMO's   {\em local calibration} procedure, adjusts the impact of the scale factors and effort
multipliers by tuning the  $a,b$ values of Equation~\ref{eq:cocII}
while keeping the other values of the tuning matrix constant as
shown in \fig{coc2}. Effectively, local calibration trims
a 23 variable model
into a model with two variables: (one  to adjust the linear effort
multipliers, and another to adjust the exponential scale factors).

Menzies' preferred local calibration procedure is the COCONUT
procedure of \fig{coconut} (first written in 2002
and first published in 2005~\cite{me04h}). 
For some number of {\em repeats},
COCONUT will {\em ASSESS} some {\em GUESSES} 
 for $(a,b)$ by applying them to some
{\em training} data. If any of these guesses prove to
be {\em useful} (i.e. reduce the estimation error) then COCONUT will recurse after
{\em constricting} the guess range for $(a,b)$ by some amount (say, by $2/3$rds). COCONUT terminates
when (a)~nothing better is found at the current level of recursion
or (b)~after 10 recursive calls-- at which point the guess range
has been constricted to  $(2/3)^{10}\approx 1$\% of the initial range.

\begin{figure}
\begin{lstlisting}
def COCONUT(training,          # list of projects
            a=10, b=1,         # initial  (a,b) guess
            deltaA    = 10,    # range of "a" guesses 
            deltaB    = 0.5,   # range of "b" guesses
            depth     = 10,     # max recursive calls
            constricting=0.66):# next time,guess less
            
  if depth > 0:
    useful,a1,b1= GUESSES(training,a,b,deltaA,deltaB)
    
    if useful: # only continue if something useful
      return COCONUT(training, 
                     a1, b1,  # our new next guess
                     deltaA * constricting,
                     deltaB * constricting,
                     depth - 1)
  return a,b

def GUESSES(training, a,b, deltaA, deltaB,
           repeats=20): # number of guesses
           
  useful, a1,b1,least,n = False, a,b, 10**32, 0
  
  while n < repeats:
    n += 1
    aGuess = a1 - deltaA + 2 * deltaA * rand()
    bGuess = b1 - deltaB + 2 * deltaB * rand()
    error  = ASSESS(training, aGuess, bGuess)
    
    if error < least: # found a new best guess
      useful,a1,b1,least = True,aGuess,bGuess,error
      
  return useful,a1,b1

def ASSESS(training, aGuess, bGuess):

   error = 0.0
   
   for project in training: # find error on training
     predicted = COCOMO2(project, aGuess, bGuess)
     actual    = effort(project)
     error    += abs(predicted - actual) / actual
     
   return error / len(training) # mean training error
\end{lstlisting}
\caption{COCONUT  tunes  $a,b$ 
of \fig{coc2}'s COCOMO function.}\label{fig:coconut}
\end{figure}

\begin{figure}
\begin{lstlisting}
def RIG():

 DATA = { COC81, NASA83, COC05, NASA10 }
 
 for data in DATA: # e.g. data = COC81
 
     errors= {}
     for learner in LEARNERS: #e.g. learner=COCONUT 
       for n in range(10): # ten times repeat
       
         for project in DATA: #  e.g.  one project
           training = data - project # leave-one-out
           model    = learn(training)
           estimate = guess(model, project)
           actual   = effort(project)
           error    = abs(actual - estimate)/actual
           errors[learner][n] = error
           
     print rank(errors) # some statistical tests
\end{lstlisting}
\caption{The experimental rig used in this paper.}\label{fig:rig}
\end{figure}

%Local calibration
%can dramatically improve the effort estimates
%from COCOMO. For example, in one result shown below, 
%COCONUT reduced the variance in the
%estimates  by a factor of seven
%(from 214\% to 34\%).

\section{Experimental Methods} 

In this section, we discuss the methods used to explore the research questions defined
in the introduction.
%% \begin{figure*}
%% {\scriptsize
%% \begin{verbatim}
%% vl=1; l=2; n=3; h=4; vh=5; xh=6

%% def nasa93():  
%%   return dict( 
%%     names= [ 
%%      'Prec','Flex','Resl','Team','Pmat',  # scale factors
%%      'rely','data','cplx','ruse','docu',  # effort multipliers
%%      'time','stor','pvol','acap','pcap',  # effort multipliers
%%      'pcon','aexp','plex','ltex','tool',  # effort multipliers
%%      'site', 'sced',                      # effort multipliers
%%      'kloc','effort],

%%     projects=[                                      
%%      #Scale        
%%      #factors    Effort multipliers                Kloc   Effort
%%      #---------- --------------------------------- -----  ------
%%      [h,h,h,vh,h,h,l,h,n,n,n,n,l,n,n,n,n,n,h,n,n,l, 25.9, 117.6],
%%      [h,h,h,vh,h,h,l,h,n,n,n,n,l,n,n,n,n,n,h,n,n,l, 24.6, 117.6],
%%      [h,h,h,vh,h,h,l,h,n,n,n,n,l,n,n,n,n,n,h,n,n,l,  7.7,  31.2],
%%      [h,h,h,vh,h,h,l,h,n,n,n,n,l,n,n,n,n,n,h,n,n,l,  8.2,  36  ],
%%      # ... 
%%     ])
%% \end{verbatim}}
%% \caption{Sample data used in this project (first four rows of NASA93).}\label{fig:data1}
%% \end{figure*}

\subsection{Choice of Experimental Rig}

``Ecological inference''
is the conceit 
that ``what holds for all, also holds for parts 
of the population"~\cite{posnet11,me12d}.
To avoid ecological inference,
our  rig in Figure~\ref{fig:rig}
runs separately for each data set.

Since some of our methods include a stochastic
algorithm (the COCONUT algorithm of \fig{coconut}),
we repeat our experimental rig   $N=10$ times
(10 was selected since, after experimentation, we
found our results looked the same at $N=8$ and
$N=16$).

\begin{changed}
It is important to note that Figure~\ref{fig:rig} is a ``leave-one-out experiment''; i.e.
training is conducted on all-but-one example, then tested
on the ``holdout'' example not seen in training. This separation of training and testing
data is of particular
importance in this study. 
As shown in \fig{types}, our  data sets (NASA10, COC81, NASA93, and COC05)
contain information on 17, 63, 92, and 93  projects, respectively. When fitted to
the   24 parameters of the standard COCOMO model  (shown in \fig{cparems}),
there may not be enough information to constrain the learning-- which means that it is theoretically
possible that data could be fitted to almost anything (including {\em spurious noise}).
To detect such spurious models, it is vital to test the learned model against some
outside source such as the holdout example.
\end{changed} 

We assess  
performance via {\em Standardized Error}($SE$); i.e.
% \begin{equation}\label{eq:mre}
% \mbox{$ \mathit{MRE}=\frac{abs(\mathit{actual} - \mathit{predicted})}{\mathit{actual}}$}.
% \end{equation}
\begin{equation}\label{eq:se}
\mbox{$ \mathit{SE}=\frac{\sum_{i=1}^{n}(abs(\mathit{actual_i} - \mathit{predicted_i}))/n}{\sum_{i=1}^{n}(abs(\mathit{actual_i} - \mathit{sampled}))/n}$} * 100
\end{equation}
This measure is derived from {\em Standardized Accuracy}($SA$) defined by Shepperd \& MacDonnell~\cite{shepperd12a} ($SE = 1 - SA$). In Equation ~\ref{eq:se}, $actual$ represents the true value, $predicted$ represents the estimated value by the predictor and $sampled$ is a value drawn randomly from a list of random samples(with replacement) from the training data.  Shepperd and MacDonnell also propose
another measure that reports the performance as a
ratio of some other, much
simpler, ``straw man'' approach (they recommend the
mean effort value of $N>100$ random samples of the
training data).

\subsection{Choice of Learners}\label{sect:whatlearn}

Our LOC(n) ``straw man'' 
estimators just uses  lines of code
in the $n$ nearest projects. For distance,
we use:
\begin{equation}\label{eq:dist}
\mathit{dist}(x,y) = \sqrt{\sum_i w_i (x_i-y_i)^2}
\end{equation} 
where $x_i,y_i$ 
are values normalized 0..1 for the range min..max
and $w_i$ is a weighting factor (defaults to $w_i=1$).
When  estimating for $n>1$ neighbors,
we combine estimates via the triangle 
function of  Walkerden
and Jeffery~\cite{Walkerden1999}; 
e.g.. for $loc(3)$, the  estimate
from the first, second and third closest neighbor with estimates
$a,b$ and $c$ respectively is
\HERE{Reviewer1c}
\begin{equation}\label{eq:tri}
\mathit{effort} = (3a + 2b + 1c)/6
\end{equation}

\begin{BLUE}
  \HERE{Reviewer2c}
We also baseline the COCOMO-II and COCONUT models using the \textit{Automatically Transformed Linear Baseline Model}(ATLM) proposed by Whigham et al. \cite{whigham15}. ATLM is a multiple linear regression model of the form

\begin{equation}
    \label{eq:atlm}
    \mathit{effort}_i = \beta_0 + \beta_1.x_{1i} + \beta_2.x_{2i} + \ldots + \beta_n.x_{ni} + \epsilon_i
\end{equation}

where $effort_i$ is the quantitative response(effort) for project $i$ and $x_i$ are the independent variables of describing the project. The prediction weights $\beta_i$ are determined using a least square error estimation. Transformations are also employed on the independent variables($x_i$) based on their nature. If the variable is continuous in nature, either a logarithmic, a square root transformation or no transformation is employed such that the skewness of the independent variable in the training set is minimized. If the variable is of categorical nature, no transformation is performed on the model.

\end{BLUE}

Apart from the LOC ``straw man'' and the ATLM baseline
we also compare COCOMO-II and COCONUT with CART
and Knear(n) as they proved their value  in the 1990s~\cite{shepperd97,Walkerden1999}. That said, CART and Knear(n)
still have currency: 
recent results from IEEE TSE 2008 and 2012 still endorse their  use for effort estimation~\cite{dejaeger12,koc11a,keung2008b}).
Also, according to the Shaw's timetable for industry adoption of research innovations
(discussed in the introduction),  CART and Knear(n) should now be mature enough for industrial use.
Further, to account for some of the more recent work on effort estimation, we also use TEAK and PEEKING2~\cite{koc11b,papa13}.

CART~\cite{breiman84} is an {\em iterative dichotomization} algorithm
that finds the attribute that most divides the data such that
the variance of the goal variable in each division is minimized.
The algorithm then recurses on each division. 
Finally, the cost data in the leaf divisions
are averaged to generate the estimate.

Knear(n) estimates a new project's effort
by a nearest neighbor  method~\cite{shepperd97}. Unlike LOC(n),
a Knear(n) method uses all attributes
(all scale factors and effort multipliers as well as lines of code)
to find the {\em n-th} nearest projects in the training data. 
Knear(3) combines efforts from three nearest neighbors using
Equation~\ref{eq:tri}.
Knear(n) is an example of CBR; i.e.  {\em case-based reasoning}.
CBR for effort estimation was 
first pioneered by Shepperd \& Schofield
in 1997~\cite{shepperd97}.
  Since then, it 
has been used extensively in software effort
estimation~\cite{Auer2006,Walkerden1999,%
  Kirsopp2002,shepperd97,kadoda00,Li2008,Li2006,Li2007,Li2009a,
  keung2008a,keung2008b,keung2008c}.  
There are several reasons  for this. Firstly, 
it works even if the domain data is sparse~\cite{Myrtveit}.
Secondly, 
unlike other predictors, it makes no assumptions about data
distributions or some  underlying parametric model. 

TEAK is built
on the assumption that spurious noise leads to large variance in the recorded efforts~\cite{koc11b}.
TEAK's pre-processor removes such regions of high variance as follows.
First, it  applies greedy agglomerate clustering  to generate a tree of clusters.
Next, it reflects on the variance
of the efforts seen in each sub-tree and discards the sub-trees with largest variance. Estimation is then performed
on the surviving examples.
PEEKING2~\cite{papa13} is a far more aggressive ``data pruner'' than TEAK and combines   data reduction operators, 
feature weighting, and Principal Component Analysis(PCA). PEEKING2 is described in  \fig{peeking}.
\begin{BLUE}\HERE{Reviewer2d}
  One important detail with TEAK and PEEKING2 is that when they prune data, they
  only do so on the {\em training} data. Given a test set, TEAK and PEEKING2 will always try to
  generate estimates for all members of that test set.
  \end{BLUE}

\begin{figure}[!t]
\small
\begin{tabular}{|p{.95\linewidth}|}\hline
\bi
\item
PEEKING2's feature weighting scheme changes  $w_i$ in \eq{dist}  according to how much an attribute
can divide and reduce the variance of the effort data (the {\em greater} the reduction, the
{\em larger} the $w_i$ score).  
\item
PEEKING2's PCA tool uses an accelerated   principle component analysis that synthesises  new
attributes $e_i, e_2,...$
that extends across the dimension of greatest  variance in the data  with attributes $d$.  
PCA  combines
redundant  variables into a smaller set of variables  (so $e \ll d$) since those
redundancies become (approximately) parallel lines
in $e$ space. For all such redundancies \mbox{$i,j \in d$}, we 
can ignore $j$ 
since effects that change over $j$ also
change in the same way over $i$.
PCA is also useful for skipping over noisy variables from $d$-- these
variables are effectively ignored since    they  do not contribute to the variance in the data.
\item
PEEKING2's prototype generator  clusters the data along the dimensions
found by accelerated PCA. Each cluster is then replaced with a ``prototype'' generated from
the median value of all attributes in that cluster. Prototype generation is a useful tool for
handling outliers: large groups of outliers get their own cluster; small sets of outliers
get ignored via median prototype generation.
\item
PEEKING2 generates estimates for a test case by finding its nearest cluster,
then the two nearest neighbors within that cluster  (where ``near''
is computed using \eq{dist} plus feature weighting). If these neighbors are found at distance
$n_1,n_2, n_1 < n_2$ and have effort values $E_1,E_2$ then the final estimate is an extrapolation
favoring the closest one:
\ei
\[
n=n_i+n_2;\;\mathit{estimate}=E_1\frac{n_2}{n} + E_2\frac{n_1}{n}
\]\\\hline
\end{tabular} 
\caption{Inside PEEKING2~\cite{papa13}.}\label{fig:peeking}
\end{figure}

\subsection{Choice of Statistical Ranking Methods}\label{sect:stats}
The last line of our experimental rig shown in
\fig{rig} {\em rank}s multiple methods for learning
effort estimators.
For this paper, those multiple methods
are the range of $l$ treatments of size
$\mathit{ls}=|\;l\;|$ explored within
each research question. For example, {\em RQ1}
studies the differences in output produced by $\mathit{ls}=4$ methods: two COCOMO variants and two others
that just use lines of code counts.

This study ranks methods using the Scott-Knott
procedure recommended by Mittas \& Angelis in their 2013
IEEE TSE paper~\cite{mittas13}.  This method
sorts a list of $l$ treatments with $\mathit{ls}$ measurements by their median
score. It then
splits $l$ into sub-lists $m,n$ in order to maximize the expected value of
 differences  in the observed performances
before and after divisions. For example,
for {\bf RQ1}, we would sort $\mathit{ls}=4$ 
methods based on their median score,
then divide them into three sub-lists of of size $\mathit{ms},\mathit{ns} \in \{(1,3), (2,2), (3,1)\}$.
Scott-Knott would declare one of these divisions
to be ``best'' as follows.
For lists $l,m,n$ of size $\mathit{ls},\mathit{ms},\mathit{ns}$ where $l=m\cup n$, the ``best'' division maximizes $E(\Delta)$; i.e.
the difference in the expected mean value
before and after the spit: 
 \[E(\Delta)=\frac{ms}{ls}abs(m.\mu - l.\mu)^2 + \frac{ns}{ls}abs(n.\mu - l.\mu)^2\]
Scott-Knott then checks if that
``best'' division is actually useful.
To implement that check, Scott-Knott would
apply some statistical hypothesis test $H$ to check
if $m,n$ are significantly different. If so, Scott-Knott then recurses on each half of the ``best'' division.

For a more specific example, consider the results
from $l=5$ treatments:

{\small \begin{verbatim}
        rx1 = [0.34, 0.49, 0.51, 0.6]
        rx2 = [0.6,  0.7,  0.8,  0.9]
        rx3 = [0.15, 0.25, 0.4,  0.35]
        rx4=  [0.6,  0.7,  0.8,  0.9]
        rx5=  [0.1,  0.2,  0.3,  0.4]
\end{verbatim}}
\noindent
After sorting and division, Scott-Knott declares:
\bi
\item Ranked \#1 is rx5 with median= 0.25
\item Ranked \#1 is rx3 with median= 0.3
\item Ranked \#2 is rx1 with median= 0.5
\item Ranked \#3 is rx2 with median= 0.75
\item Ranked \#3 is rx4 with median= 0.75
\ei
Note that Scott-Knott found  little
difference between rx5 and rx3. Hence,
they have the same rank, even though their medians differ.

Scott-Knott is better than an 
 all-pairs hypothesis test of all methods; e.g. six treatments
can be compared \mbox{$(6^2-6)/2=15$} ways. 
A 95\% confidence test run for each comparison has  a very low total confidence: 
\mbox{$0.95^{15} = 46$}\%.
To avoid an all-pairs comparison, Scott-Knott only calls on hypothesis
tests {\em after} it has found splits that maximize the performance differences.
 
For this study, our hypothesis test $H$ was a
conjunction of the A12 effect size test of  and
non-parametric bootstrap sampling; i.e. our
Scott-Knott divided the data if {\em both}
bootstrapping and an effect size test agreed that
the division was statistically significant (99\%
confidence) and not a ``small'' effect ($A12 \ge
0.6$).

For a justification of the use of non-parametric
bootstrapping, see Efron \&
Tibshirani~\cite[p220-223]{efron93}.
For a justification of the use of effect size tests
see Shepperd \& MacDonell~\cite{shepperd12a}; Kampenes~\cite{kampenes07}; and
Kocaguneli et al.~\cite{kocharm13}. These researchers
warn that even if an
hypothesis test declares two populations to be
``significantly'' different, then that result is
misleading if the ``effect size'' is very small.
Hence, to assess 
the performance differences 
we first must rule out small effects.
Vargha and Delaney's
non-parametric 
A12 effect size test 
explores
two lists $M$ and $N$ of size $m$ and $n$:
\[A12 = \left(\sum_{x\in M, y \in N} 
\begin{cases} 
1   & \mathit{if}\; x > y\\
0.5 & \mathit{if}\; x == y
\end{cases}\right) / (mn)
\]
This expression computes the probability that numbers in one sample are bigger than in another.
This test was recently 
endorsed by Arcuri and Briand
at ICSE'11~\cite{arcuri11}.

\section{Results}
\subsection{COCOMO vs Just Lines of Code}\label{sect:justloc}
This section explores {\bf RQ1:
is parametric estimation no better than 
using simple lines of code measures?}

An often heard, but not often tested, criticism of parametric
estimation methods is that they are no
better than just using simple lines of code measures.
As shown in \fig{loc}, this is not necessarily true.
This figure is a comparative ranking for LOC(1)
LOC(3), COCOMO-II and COCONUT.
The rows of \fig{loc} are sorted by the SE figures.
These rows are divided according to their 
 {\em rank}, shown in the left column: better methods
have {\em lower rank} since they have {\em lower SE} error values.
The right-hand-side column displays the median error (as a black dot)
inside the inter-quartile range
(25th to 75th percentile, show as a horizontal line).

The key feature of \fig{loc}
is that  just using lines of
code is {\em not}  better than parametric estimation.
If the reader is surprised by this result, then we note that
with a little
 mathematics, it is possible to show
 that the results of \fig{loc} are not surprising.
 From \eq{cocII}, recall that
the minimum  
effort  is bounded by the  {\em sum} of the minimum scale factors
and the {\em product} of the minimum effort multipliers.
Similar expressions hold for the  maximum effort estimate. Hence,
for a given KLOC, the range of values is given by:
\[
0.18*\mathit{KLOC}^{0.97}  \le \mathit{effort} \le 154*\mathit{KLOC}^{1.23}\]
Dividing the minimum and maximum values results in an  expression showing
how    effort can vary for any given KLOC.: 
\begin{equation}\label{eq:ration}
154/0.18 *\mathit{KLOC}^{1.23 - 0.97} = 856*\mathit{KLOC}^{0.25}
\end{equation} 
 \eq{ration} explains why just using KLOC performs so badly. 
That equation had two components: KLOC raised to
a small exponent (0.25), and a constant showing the influence of all  other
COCOMO variables. The large value of 856 for that second component
indicates that many factors outside of KLOC influence effort. Hence, it is hardly
surprising that just using KLOC is a poor way to do effort estimation.

\begin{figure}[!t]
 
{\small
{\bf NASA10 (new NASA data up to 2010):}

{\small \begin{tabular}{l@{~~~}l@{~~~}r@{~~~}r@{~~~}c}
\arrayrulecolor{darkgray}
\rowcolor[gray]{.9}  rank & treatment & median & IQR & 
%min= 20, max= 117
\\
  1 &      COCOMO-II &    37  &  57 & \quart{0}{57}{37}{82} \\
 1 &      COCONUT &    39  &  54 & \quart{0}{54}{39}{82} \\
 1 &       loc(3) &    47  &  93 & \quart{10}{93}{47}{82} \\
  1 &       loc(1) &    75  &  98 & \quart{0}{98}{75}{82} \\
\end{tabular}}

% :learn 4.64 :analyze 1.69 :boots 3 effects 5 :conf 0.970299

~\\

{\bf COC05 (new COCOMO data up to 2005):}

{\small \begin{tabular}{l@{~~~}l@{~~~}r@{~~~}r@{~~~}c}
\arrayrulecolor{darkgray}
\rowcolor[gray]{.9}  rank & treatment & median & IQR & \\%min= 20, max= 166\\
  1 &      COCOMO-II &    12  &  52 & \quart{0}{52}{12}{54} \\
  \hline 2 &       loc(1) &    21  &  56 & \quart{0}{56}{21}{54} \\
  2 &       loc(3) &    22  &  55 & \quart{0}{55}{22}{54} \\
  2 &      COCONUT &    22  &  89 & \quart{0}{89}{22}{54} \\
\end{tabular}}

% :learn 763.316602 :analyze 6.61469 :boots 1 effects 1 :conf 0.99
%\subsection{xyz14deTune}

% :learn 13.41 :analyze 2.49 :boots 2 effects 2 :conf 0.9801

~\\

{\bf NASA93 (NASA data up to 1993):}

{\small \begin{tabular}{l@{~~~}l@{~~~}r@{~~~}r@{~~~}c}
\arrayrulecolor{darkgray}
\rowcolor[gray]{.9}  rank & treatment & median & IQR & 
%min= 15, max= 129
\\
  1 &      COCONUT &    12  &  48 & \quart{0}{48}{12}{74} \\
  1 &      COCOMO-II &    15  &  50 & \quart{0}{50}{15}{74} \\
\hline  2 &       loc(1) &    23  &  63 & \quart{0}{63}{23}{74} \\
  2 &       loc(3) &    35  &  65 & \quart{0}{65}{35}{74} \\
\end{tabular}}

 %:learn 142.78 :analyze 10.13 :boots 3 effects 11 :conf 0.970299

~\\

{\bf COC81 (original data from the 1981 COCOMO book):}

{\small \begin{tabular}{l@{~~~}l@{~~~}r@{~~~}r@{~~~}c}
\arrayrulecolor{darkgray}
\rowcolor[gray]{.9}  rank & treatment & median & IQR & %min= 14, max= 291
\\
  1 &      COCOMO-II &    3  &  21 & \quart{0}{21}{3}{31} \\
  1 &      COCONUT &    4  &  24 & \quart{0}{24}{4}{31} \\
\hline  2 &       loc(3) &    14  &  36 & \quart{0}{36}{14}{31} \\
  2 &       loc(1) &    19  &  42 & \quart{0}{42}{19}{31} \\
\end{tabular}}

% :learn 63.13 :analyze 6.36 :boots 3 effects 8 :conf 0.970299

}
\caption{COCOMO vs just lines
of code. SE values seen in 
leave-one-studies, repeated ten times.
For each of the four tables in this figure,
{\em better} methods appear {\em higher} in the tables.
In these tables,
median and IQR are the 50th and the 
(75-25)th percentiles. The IQR range is
shown  in the right column
with black dot at the median. Horizontal lines
divide the ``ranks'' found by our Scott-Knott+bootstrapping+effect size tests  (shown in  left column).
}\label{fig:loc}
\end{figure}

\begin{figure}[!b] 
{\small

~\\

\noindent {\bf NASA10: (new NASA data up to 2010):}

{\small \begin{tabular}{l@{~~~}l@{~~~}r@{~~~}r@{~~~}c}
\arrayrulecolor{darkgray}
\rowcolor[gray]{.9}  rank & treatment & median & IQR & \\%min= 24, max= 104\\
  1 &      COCOMO-II &    34  &  55 & \quart{10}{65}{34}{94} \\
  1 &      COCONUT &    41  &  61 & \quart{0}{61}{41}{94} \\
  1 &     CART &    46  &  55 & \quart{10}{65}{46}{94} \\
  1 &     Knear(1) &    49  &  89 & \quart{9}{98}{49}{94} \\
\hline  2 &   Knear(3)   &    71  &  104 & \quart{0}{104}{71}{94} \\
\hline 3 & ATLM &    90  &  77 & \quart{0}{90}{77}{94} \\ 
\end{tabular}}
~\\

\noindent
{\bf COC05: (new COCOMO data up to 2005):}

{\small \begin{tabular}{l@{~~~}l@{~~~}r@{~~~}r@{~~~}c}
\arrayrulecolor{darkgray}
\rowcolor[gray]{.9}  rank & treatment & median & IQR & %min= 18, max= 166\\
\\
  1 &   COCOMO-II &    13  &  51 & \quart{0}{51}{13}{55} \\
  1 &   CART &    14  &  48 & \quart{0}{48}{14}{55} \\
\hline  2 &     Knear(1) &    22  &  51 & \quart{0}{51}{22}{55} \\
  2 &     Knear(3) &    22  &  54 & \quart{0}{54}{22}{55} \\
  2 &      COCONUT &    22  &  81 & \quart{0}{81}{22}{55} \\ 
\hline 3 &   ATLM &    94  &  47 & \quart{49}{47}{94}{55} \\ 
\end{tabular}}

~\\

\noindent {\bf NASA93: (NASA data up to 1993):}

{\small \begin{tabular}{l@{~~~}l@{~~~}r@{~~~}r@{~~~}c}
\arrayrulecolor{darkgray}
\rowcolor[gray]{.9}  rank & treatment & median & IQR & \\%min= 15, max= 110\\
  1 &      COCONUT &    13  &  48 & \quart{0}{48}{13}{89} \\
  1 &      COCOMO-II &    15  &  50 & \quart{0}{50}{15}{89} \\
\hline  2 & Knear(1) &    33  &  71 & \quart{0}{71}{33}{89} \\ 
  2 &     Knear(3) &    34  &  63 & \quart{0}{63}{34}{89} \\
  2 &     CART &    34  &  63 & \quart{0}{63}{34}{89} \\
\hline 3 & ATLM &    53  &  56 & \quart{25}{81}{56}{94} \\
\end{tabular}}

~\\

\noindent {\bf COC81: (original data from the 1981 COCOMO book):}

{\small \begin{tabular}{l@{~~~}l@{~~~}r@{~~~}r@{~~~}c}
\arrayrulecolor{darkgray}
\rowcolor[gray]{.9}  rank & treatment & median & IQR & \\%min= 14, max= 306\\
  1 &      COCOMO-II &    3  &  20 & \quart{0}{20}{3}{29} \\
  1 &      COCONUT &    4  &  25 & \quart{0}{25}{4}{29} \\
\hline  2 &         CART &    13  &  37 & \quart{0}{37}{13}{29} \\
 2 &     Knear(3) &    19  &  48 & \quart{0}{48}{19}{29} \\
\hline 3 &     Knear(1) &    30  &  75 & \quart{0}{75}{30}{29}\\ 
\hline 4 &          ATLM &    75  &  42 & \quart{35}{42}{75}{94} \\
\end{tabular}}}

\caption{COCOMO vs standard methods.
Displayed as per \fig{loc}. }\label{fig:standard}
\end{figure}

\subsection{COCOMO vs Other Methods}\label{sect:othermethods}
This section explores {\bf RQ2: 
Has parametric estimation been superseded
by more recent estimation methods?}
and {\bf RQ3: Are the old parametric tunings irrelevant to
more recent projects?}

\fig{standard} compares COCOMO and COCONUT with  
standard effort estimation methods
from the 1990s (CART and Knear(n))
\begin{BLUE}\HERE{Reviewer2f}
  as well as  ATLM (the baseline effort estimation method
  proposed in 2015 by Whigham et.al.~\cite{whigham15} (a method defined
  by its authors to better define effort
  estimation experiments-- and perhaps
  to encourage more repeatability in these
  kinds of studies).
  \end{BLUE}
In that comparison,   COCOMO-II's error is not ranked worse than any other method
(sometimes
  COCONUT had a slightly lower median SE but that difference was small: $\le 2$\%).

\fig{newer} compares COCOMO and COCONUT to more recent effort estimation methods (TEAK and PEEKING2). Once again,   nothing was ever ranked better than COCOMO-II or COCONUT.

From these results,
we recommend that effort estimation researchers take care to benchmark
their new method against older ones.

As to COCONUT, this method
was often ranked equaled to COCOMO-II.  
In several cases  COCOMO-II and COCONUT were ranked first and second and
the median difference in their scores is very small.

From this data, we conclude that it is not
always true the parametric estimation has been
superseded by more recent innovations such
as CART, Knear(n), TEAK or PEEKING2. Also, the COCOMO-II tunings from 2000
are useful not just for the projects prior to 2000
(all of COC81, plus some of NASA93)
but also for projects completed up to a decade after
those tunings (NASA10).

\begin{figure}
{\small
{\bf NASA10 (new NASA data up to 2010):}

{\small \begin{tabular}{l@{~~~}l@{~~~}r@{~~~}r@{~~~}c}
\arrayrulecolor{darkgray}
\rowcolor[gray]{.9}  rank & treatment & median & IQR & %min= 0, max= 185\\
\\
  1 &      COCONUT &    37  &  58 & \quart{0}{58}{37}{111} \\
  1 &   COCOMO-II &    38  &  56 & \quart{0}{56}{38}{111} \\
\hline 
  2 &      TEAK &    87  &  118 & \quart{20}{85}{87}{111} \\
  2 & PEEKING2 &    100  &  67 & \quart{33}{67}{100}{111}  
\end{tabular}}

~\\

{\bf COC05 (new COCOMO data up to 2005):}

{\small \begin{tabular}{l@{~~~}l@{~~~}r@{~~~}r@{~~~}c}
\arrayrulecolor{darkgray}
\rowcolor[gray]{.9}  rank & treatment & median & IQR & \\%min= 20, max= 300\\
  1 &      COCOMO-II &    13  &  55 & \quart{0}{55}{13}{110} \\
  1 & COCONUT &    20  &  86 & \quart{0}{86}{20}{110} \\ \hline
  2 &      TEAK &    33  &  84 & \quart{0}{84}{33}{110} \\ 
  2 & PEEKING2 &    34  &  79 & \quart{0}{79}{34}{110}  
\end{tabular}}

% :learn 70.71 :analyze 4.04 :boots 2 effects 2 :conf 0.9801

~\\

{\bf NASA93 (NASA data up to 1993):}

{\small \begin{tabular}{l@{~~~}l@{~~~}r@{~~~}r@{~~~}c}
\arrayrulecolor{darkgray}
\rowcolor[gray]{.9}  rank & treatment & median & IQR & %min= 15, max= 250\\
\\
  1 &    COCONUT &    12  &  50 & \quart{0}{50}{12}{100} \\
  1 &   COCOMO-II &    15  &  49 & \quart{0}{49}{15}{100}\\\hline
  1 & TEAK &    37  &  84 & \quart{0}{84}{37}{100} \\
  2 & PEEKING2 &    43  &  76 & \quart{0}{76}{43}{100}  
\end{tabular}}

% :learn 548.2 :analyze 7.81 :boots 3 effects 14 :conf 0.970299

% :learn 37.86 :analyze 3.84 :boots 3 effects 5 :conf 0.970299
%\subsection{coc81}

~\\

{\bf COC81 (original data from the 1981 COCOMO book):}

{\small \begin{tabular}{l@{~~~}l@{~~~}r@{~~~}r@{~~~}c}
\arrayrulecolor{darkgray}
\rowcolor[gray]{.9}  rank & treatment & median & IQR & %min= 12, max= 500\\
\\
  1 &      COCOMO-II &    3  &  21 & \quart{0}{21}{3}{100} \\
  1 &      COCONUT &    4  &  24 & \quart{0}{24}{4}{100} \\
\hline  2 & TEAK &    15  &  61 & \quart{0}{61}{15}{100} \\
  2 & PEEKING2 &    19  &  58 & \quart{0}{58}{19}{100} \ 
\end{tabular}}

% :learn 260.57 :analyze 8.89 :boots 6 effects 11 :conf 0.941480149401

}
\caption{COCOMO vs newer methods. Displayed as per \fig{loc}.}\label{fig:newer}
\end{figure}

\subsection{COCOMO vs Simpler COCOMO}\label{sect:simpler}
This section explores {\bf RQ4:
Is parametric estimation expensive to deploy
at some new site?}. To that end,
we assess the impact
a certain simplifications imposed onto COCOMO-II.

\subsubsection{Range Reductions}
The cost with deploying COCOMO in a new
organization is the training effort required to generate consistent project
rankings from different analysts. If we could reduce 
the current six
point scoring scale (very low, low, nominal, high, very high and extremely high)
then there would be less scope 
to disagree about projects. 
Accordingly,  we tried
reducing the  six point scale to just three:
\bi
\item {\em Nominal}: same as before;
\item {\em Above}: anything above nominal;
\item {\em Below}: anything below nominal.
\ei
To do  this, the tunings table of
\fig{coc2} was altered. For each row, all values
below nominal were replaced with their mean (and
similarly with above-nominal values).  For example,
here are the tunings for {\em time} before and after
being reduced to {\em below, nominal, above}:

{\small   
~~~~~~\begin{tabular}{r|ll|l|lll|}
      range      & vlow&  low&{\em nominal}&high&vhigh&xhigh\\\hline
     before & 1.22& 1.09& 1.00& 0.93& 0.86& 0.80\\
     reduced&1.15& 1.15& 1.00&  0.863& 0.863&0.863\\\cline{2-3}\cline{5-7}
                 & \multicolumn{2}{c|}{{\em below}} &&\multicolumn{3}{c|}{{\em above}}
\end{tabular}
 }
\subsubsection{Row Reductions}\label{sect:row}
New COCOMO models are tuned only after collecting
100s of new examples. If that was not necessary, we could look forward to multiple
COCOMO models, each tuned to different specialized (and small) samples of projects.
Accordingly, we explore tuning COCOMO
on very small data sets.

To implement  row reduction, training data was
shuffled at random and training was conducted on
all rows or  just the first four or eight  rows
(denoted {\em r4,r8} respectively). Note that, given  the positive
results obtained with {\em r8} we did not explore larger training sets.

\subsubsection{Column Reduction}\label{sect:pruner}

Prior results tell us that row reduction should be
accompanied by column reduction.  A study by Chen et
al.~\cite{chen05a} combines column reduction (that
discards noisy or correlated attributes) with row
reduction. Their results are very clear: as the
number of rows shrink, {\em better} estimates come
from using {\em fewer}
columns. Miller~\cite{miller02} explains  why this is so:  the variance of a
linear model learned by minimizing least-squares error decreases as the number of columns in the model
decreases. That is, as the number of columns decrease,
prediction reliability can increase (caveat: 
if you remove too much,
there is no information left for predictions).

Accordingly, this experiment sorts the attributes in the training set according
to how well they select for specific effort values. 
Let $x\in a_i$ denote the list of unique values seen for attribute $a_i$. Further,
let there be $N$ rows in the training data; 
let  $r(x)$ denote the $n$ rows containing $x$; and let $v(r(x))$ be the variance
of the effort value in those rows. The values of ``good'' attributes
select most for specific efforts; i.e. those attributes minimize
$E(\sigma,a_i) =\sum_{x\in a_i} \left(n/N * v(r(x))\right)$

This experiment sorted all training data attributes by $E(\sigma,a_i)$ then kept
the data in the {\em lower quarter} or  {\em half} or {\em all} of  the columns
(denoted {\em c0.25} or {\em c0.5} or {\em c1} respectively).
Note that, due to the results of \fig{loc}, LOC was excluded from column reduction.

\subsubsection{Results}

\fig{fss} compares results found when either
{\em all} or some {\em reduced} set of ranges, rows,
and columns are used. Note our nomenclature:  the
COCONUT:c0.5,r8  results are those
seen after training on eight randomly selected
training examples reduced to {\em below, nominal,
above}, while ignoring 50\% of the columns.

 \fig{fss} suggests that it is defensible
to learn a COCOMO model from just four to eight projects. Most of the
{\em r8} results are top-ranked with the exception
of the COC81 results (but even there, the absolute
difference between the top {\em r8} results and
standard COCOMO is very small: just 2\%).

Overall, \fig{fss} suggests that the modeling
effort associated with COCOMO-II could be reduced. Hence,
it need not be expensive to deploy parametric estimation
at some new site.
Projects attributes
do not need to be specified in great detail:
a simple three point scale will suffice:
 {\em below, nominal, above}. As to how much data is
required for modeling, 
a mere four to eight projects can
suffice for calibration.
Hence, it
should be possible to quickly build many COCOMO-like
models for various specialized sub-groups using just
a three-point scale

\begin{figure} 

  ~\\
  
{\small
{\bf NASA10 (new NASA data up to 2010):}

{\small \begin{tabular}{l@{~~~}l@{~~~}r@{~~~}r@{~~~}c}
\arrayrulecolor{darkgray}
\rowcolor[gray]{.9}  rank & treatment & median & IQR & %min= 15, max= 77\\
\\
  1 &      COCOMO-II &    34  &  56 & \quart{10}{56}{34}{137} \\
  1 &      COCONUT &    39  &  58 & \quart{0}{58}{39}{137} \\
  1 & COCONUT:c1,r8 &    40  &  48 & \quart{0}{48}{40}{137} \\
  1 & COCONUT:c0.5,r8 &    41  &  75 & \quart{0}{75}{41}{137} \\
\hline2 & COCONUT:c1,r4 &    47  &  59 & \quart{15}{59}{47}{137} \\
  2 & COCONUT:c0.25,r8 &    50  &  75 & \quart{12}{75}{50}{137} \\
  2 & COCONUT:c0.5,r4 &    59  &  65 & \quart{0}{65}{59}{137} \\
\hline3 & COCONUT:c0.25,r4 &    71  &  44 & \quart{30}{44}{71}{137} \\
\end{tabular}}

~\\

{\bf COC05 (new COCOMO data up to 2005):}

{\small \begin{tabular}{l@{~~~}l@{~~~}r@{~~~}r@{~~~}c}
\arrayrulecolor{darkgray}
\rowcolor[gray]{.9}  rank & treatment & median & IQR & \\%min= 20, max= 166\\
  1 &      COCOMO-II &    13  &  51 & \quart{0}{51}{13}{54} \\
\hline2 & COCONUT:c1,r8 &    19  &  53 & \quart{0}{53}{19}{54} \\
  2 & COCONUT:c0.5,r8 &    22  &  57 & \quart{0}{57}{22}{54} \\
\hline2 & COCONUT:c1,r4 &    23  &  78 & \quart{0}{78}{23}{54} \\
  2 &      COCONUT &    23  &  82 & \quart{0}{82}{23}{54} \\
  2 & COCONUT:c0.5,r4 &    24  &  68 & \quart{0}{68}{24}{54} \\
  2 & COCONUT:c0.25,r4 &    26  &  73 & \quart{0}{73}{26}{54} \\
  2 & COCONUT:c0.25,r8 &    27  &  72 & \quart{0}{72}{27}{54} \\
\end{tabular}}

% :learn 70.71 :analyze 4.04 :boots 2 effects 2 :conf 0.9801

~\\

{\bf NASA93 (NASA data up to 1993):}

{\small \begin{tabular}{l@{~~~}l@{~~~}r@{~~~}r@{~~~}c}
\arrayrulecolor{darkgray}
\rowcolor[gray]{.9}  rank & treatment & median & IQR & %min= 12, max= 91\\
\\
  1 &      COCONUT &    12  &  53 & \quart{0}{53}{12}{111} \\
  1 & COCONUT:c0.5,r8 &    13  &  46 & \quart{0}{46}{13}{111} \\
  1 &      COCOMO-II &    14  &  49 & \quart{0}{49}{14}{111} \\
  1 & COCONUT:c0.25,r8 & 15  &  48 & \quart{0}{48}{15}{111} \\
\hline2 & COCONUT:c1,r8 &    17  &  61 & \quart{0}{61}{17}{111} \\
  2 & COCONUT:c1,r4 &    18  &  59 & \quart{0}{59}{18}{111} \\
  2 & COCONUT:c0.25,r4 &    20  &  44 & \quart{0}{44}{20}{111}\\
  2 & COCONUT:c0.5,r4 &    21  &  67 & \quart{0}{67}{21}{111} \\
\end{tabular}}

% :learn 548.2 :analyze 7.81 :boots 3 effects 14 :conf 0.970299

% :learn 37.86 :analyze 3.84 :boots 3 effects 5 :conf 0.970299
%\subsection{coc81}

~\\

{\bf COC81 (original data from the 1981 COCOMO book):}

{\small \begin{tabular}{l@{~~~}l@{~~~}r@{~~~}r@{~~~}c}
\arrayrulecolor{darkgray}
\rowcolor[gray]{.9}  rank & treatment & median & IQR & %min= 14, max= 117\\
\\
  1 &      COCOMO-II &    3  &  18 & \quart{0}{18}{3}{30} \\
  1 &      COCONUT &    4  &  29 & \quart{0}{29}{4}{30} \\
  1 & COCONUT:c1,r4 &    5  &  15 & \quart{0}{15}{5}{30} \\
  1 & COCONUT:c0.5,r4 &    5  &  20 & \quart{0}{20}{5}{30} \\
\hline2 & COCONUT:c0.25,r4 & 6  &  30 & \quart{0}{30}{6}{30} \\
  2 & COCONUT:c0.5,r8 &    7  &  22 & \quart{0}{22}{7}{30}\\
  2 & COCONUT:c0.25,r8 &    7  &  36 & \quart{0}{36}{7}{30} \\
  2 & COCONUT:c1,r8 &    8  &  25 & \quart{0}{25}{8}{30} \\
\end{tabular}}

% :learn 260.57 :analyze 8.89 :boots 6 effects 11 :conf 0.941480149401

}
\caption{COCOMO vs simpler COCOMO.  
Displayed as per \fig{loc}.}\label{fig:fss}
\end{figure}

\newcommand{\crule}[3][darkgray]{\textcolor{#1}{\rule{#2}{#3}}}

\newcommand{\rone}{\crule{1mm}{1.95mm}}

\newcommand{\rtwo}{\crule{1mm}{1.95mm}\hspace{0.3pt}\crule{1mm}{1.95mm}}

\newcommand{\rthree}{\crule{1mm}{1.95mm}\hspace{0.3pt}\crule{1mm}{1.95mm}\hspace{0.3pt}\crule{1mm}{1.95mm}}

\newcommand{\rfour}{\crule{1mm}{1.95mm}\hspace{0.3pt}\crule{1mm}{1.95mm}\hspace{0.3pt}\crule{1mm}{1.95mm}\hspace{0.3pt}\crule{1mm}{1.95mm}}
 
\newcommand{\rfive}{\crule{1mm}{1.95mm}\hspace{0.3pt}\crule{1mm}{1.95mm}\hspace{0.3pt}\crule{1mm}{1.95mm}\hspace{0.3pt}\crule{1mm}{1.95mm}}

\section{Threats to Validity}

%% In the future, we plaIt is reasonable to require
%% that these results be repeated on a 
%% XXX more parametric estiamtion that COCOMO.
%% COCOMO more simialrt to SEER etc than others

%% there are other biases as the analogy guys will point out.  you are only using cocomo parameters.   
%% XXX but that might be the point- that estiamtion is easy is the right data is collected, that
%% effective effort estiaton is a metter of good data colelction rather than in the subsequent analtysis.

Questions of validity arise in terms of how the
projects (data-sets) are chosen for our experiments.
While we used all the data sets that could be shared
between our team, it is not clear if our results
would generalize to other as yet unstudied
data-sets. One the other hand, in terms of the
parametric estimation literature, this is one of the most extensive
and elaborate studies yet published.

To increase external
validity, all the data used in this work is available
on-line in the PROMISE code repository. Also, our use of a leave-one-out experimental rig
plus the public availability of three of our four data sets (NASA93, COC81, NASA10)
means that other researchers would be able to reproduce exactly
our rig on exactly the code used in this study.

One source of bias in this study
are the learners used for the defect prediction
studies. Data mining is a large and active field and
any single study can only use a small subset of the
known data mining algorithms.
Any case studies
in SE data mining  can only explore a small
subset of options, selected by the biases of the researcher. The best any researcher
can hope to do is state their biases and make some attempt to compensate for them.
Accordingly:
\bi
\item
The
biases of the authors of this paper made us select
a parametric modeling method (COCOMO)  as the main modeling method.
\item
We then made a conscious decision to reverse those biases
and explore non-parametric methods (PEEKING2 and TEAK) as well
as decision-tree methods (CART).
\ei

\section{Conclusion}
The past few decades have seen a long line of innovative  methods
applied to effort estimation. This paper has compared a sample of those methods
to a decades-old parametric estimation method. 

Based on that study, we offered a negative result in which a decades
old effort estimation method performed as well, or better,
as more recent methods:
\bi
\item {\bf RQ1}: just using LOC for estimation is far worse
that parametric estimation over many attributes (see \tion{justloc}); 
\item {\bf RQ2}: new innovations in effort estimation have not superseded parametric estimation (see \tion{othermethods});
\item {\bf RQ3}: Old parametric tunings are not out-dated (see \tion{othermethods});
\item {\bf RQ4}: It is possible to simplify parametric estimation with some range, row and column pruning to reduce the cost
of deploying those methods at a new site (see \tion{simpler});
%\item {\bf RQ5}: Parametric estimation methods like COCOMO that assume effort is exp%onential on lines of code are {\em not} unduly
%sensitive to errors in the LOC measure (see \tion{nonoise});.
\ei
Hence, we conclude that in 2016, it is still a valid and a recommended practice to {\em first} try parametric estimation.
In these experiments, four to eight projects were enough to learn good predictors (and we are exploring methods
to reduce that even further). This is an important result since, given the rapid pace of change
on software engineering, it is unlikely organizations will have access to dozens and dozens of prior
relevant projects to learn from.

Our take-away message here is that the choice of data to collect may be more
important than what learner is applied to that data. Certainly, it is true that
not all projects can be expressed in terms of COCOMO. But when there is a choice,
we recommend collecting data like Figure~3, and then processing that data using COCOMO-II.

\section{Future Work}

The negative results of this paper makes us question 
some of the newer (and supposedly better) innovative techniques for effort estimation.  
The unique and highly variable characteristics of SE
project data place great limitation on the results
obtained by naively applying some brand-new
algorithm.  Perhaps one direction for future direction is to
investigate how innovative new techniques can extend
(rather than replace) existing and successful
estimation methods.

Having endorsed the use of parametric methods such as COCOMO, it is appropriate
to discuss current plans for new versions of that approach.
Recent changes  in the software industry
suggest  it is time  to revise COCOMO-II.
The rise of agile methods, web
services, cloud services, parallelized software on
multi-core chips, field-programmable-gate-array
(FPGA) software, apps, widgets, and net-centric
systems of systems (NCSOS) have caused the COCOMO II
developers and users to begin addressing an upgrade
to the 14-year-old COCOMO II. 
Current discussions
of a potential COCOMO III have led to a
reconsideration of the old COCOMO 1981 development
modes, as different development phenomena appear to
drive the costs and schedules of web-services,
business data processing, real-time embedded
software, command and control, and engineering and
scientific applications. 

Additionally, while calibrating COCOMO II model and developing COCOMO III, we were also seeing time-competitive Agile projects in well-jelled, domain-experienced rapid development organizations, which demonstrates tremendous effort reduction and schedule acceleration~\cite{ingold13}. Finally, the emerging community-based software development, i.e. software crowd sourcing~\cite{yang13}, challenges the underlying assumptions of traditional software estimation laws. Access to external workforce and competition factors are becoming critical development influential factors and need to be further investigated.   

Efforts to characterize
these models and to gather data to calibrate models
for dealing with them are underway. Contributors
to the definition and calibration are most  welcome.

\section*{Acknowledgements}
 The research described in this paper was carried out, in part, at the Jet
 Propulsion Laboratory, California Institute of Technology,
 under a contract with the US National Aeronautics and
 Space Administration. Reference herein to any specific
 commercial product, process, or service by trade name,
 trademark, manufacturer, or otherwise does not constitute
or imply its endorsement by the US Government.

% That's all folks!
\vspace*{0.5mm} 

\bibliographystyle{plain}

\bibliography{refs,refs1}

\begin{thebibliography}{10}

\bibitem{arcuri11}
A.~Arcuri and L.~Briand.
\newblock A practical guide for using statistical tests to assess randomized
  algorithms in software engineering.
\newblock In {\em ICSE'11}, pages 1--10, 2011.

\bibitem{Auer2006}
Martin Auer, Adam Trendowicz, Bernhard Graser, Ernst Haunschmid, and Stefan
  Biffl.
\newblock Optimal project feature weights in analogy-based cost estimation:
  Improvement and limitations.
\newblock {\em IEEE Trans. Softw. Eng.}, 32:83--92, 2006.

\bibitem{baker07}
Dan Baker.
\newblock A hybrid approach to expert and model-based effort estimation.
\newblock Master's thesis, Lane Department of Computer Science and Electrical
  Engineering, West Virginia University, 2007.
\newblock Available from
  \url{https://eidr.wvu.edu/etd/documentdata.eTD?documentid=5443}.

\bibitem{black77}
R.~Black, R.~Curnow, R.~Katz, and M.~Bray.
\newblock Bcs software production data, final technical report radc-tr-77-116.
\newblock Technical report, Boeing Computer Services, Inc., March 1977.

\bibitem{boehm81}
B.~Boehm.
\newblock {\em Software Engineering Economics}.
\newblock Prentice Hall, 1981.

\bibitem{boehm00a}
B.~Boehm.
\newblock Safe and simple software cost analysis.
\newblock {\em IEEE Software}, pages 14--17, September/October 2000.

\bibitem{boehm00b}
Barry Boehm, Ellis Horowitz, Ray Madachy, Donald Reifer, Bradford~K. Clark,
  Bert Steece, A.~Winsor Brown, Sunita Chulani, and Chris Abts.
\newblock {\em Software Cost Estimation with Cocomo II}.
\newblock Prentice Hall, 2000.

\bibitem{breiman84}
L.~Breiman, J.~H. Friedman, R.~A. Olshen, and C.~J. Stone.
\newblock {\em Classification and Regression Trees}.
\newblock 1984.

\bibitem{burgess01}
C.J. Burgess and Martin Lefley.
\newblock Can genetic programming improve software effort estimation? a
  comparative evaluation.
\newblock {\em Information and Software Technology}, 43(14):863--873, December
  2001.

\bibitem{chen05a}
Zhihao Chen, Barry Boehm, Tim Menzies, and Daniel Port.
\newblock Finding the right data for software cost modeling.
\newblock {\em IEEE Software}, 22:38--46, 2005.

\bibitem{chen05}
Zhihoa Chen, Tim Menzies, and Dan Port.
\newblock Feature subset selection can improve software cost estimation.
\newblock In {\em PROMISE'05}, 2005.
\newblock Available from \url{http://menzies.us/pdf/05/fsscocomo.pdf}.

\bibitem{chulani99}
S.~Chulani, B.~Boehm, and B.~Steece.
\newblock Bayesian analysis of empirical software engineering cost models.
\newblock {\em IEEE Transaction on Software Engineerining}, 25(4), July/August
  1999.

\bibitem{cohen95}
P.R. Cohen.
\newblock {\em Empirical Methods for Artificial Intelligence}.
\newblock MIT Press, 1995.

\bibitem{cora10}
A.~Corazza, S.~Di~Martino, F.~Ferrucci, C.~Gravino, F.~Sarro, and E.~Mendes.
\newblock How effective is tabu search to configure support vector regression
  for effort estimation?
\newblock In {\em Proceedings of the 6th International Conference on Predictive
  Models in Software Engineering}, PROMISE '10, pages 4:1--4:10, 2010.

\bibitem{cordero97}
R.~Cordero, M.~Costamagna, and E.~Paschetta.
\newblock A genetic algorithm approach for the calibration of cocomo-like
  models.
\newblock In {\em 12th COCOMO Forum}, 1997.

\bibitem{dabney07}
J.~B. Dabney.
\newblock Return on investment for {IV\&V}, 2002-2004.
\newblock NASA funded study. Results Available from
  \url{http://sarpresults.ivv.nasa.gov/ViewResearch/24.jsp}.

\bibitem{dejaeger12}
Karel Dejaeger, Wouter Verbeke, David Martens, and Bart Baesens.
\newblock Data mining techniques for software effort estimation: A comparative
  study.
\newblock {\em IEEE Transactions on Software Engineering}, 38:375--397, 2012.

\bibitem{efron93}
Bradley Efron and Robert~J Tibshirani.
\newblock {\em An introduction to the bootstrap}.
\newblock Mono. Stat. Appl. Probab. Chapman and Hall, London, 1993.

\bibitem{frei79}
F.~Freiman and R.~Park.
\newblock Price software model - version 3: An overview.
\newblock In {\em Proceedings, IEEE-PINY Workshop on Quantitative Software
  Models, IEEE Catalog Number TH 0067-9}, pages 32--41, October 1979.

\bibitem{herd77}
J.~Herd, J.~Postak, W.~Russell, and J.~Stewart.
\newblock Software cost estimation study-study results, final technical report,
  radc-tr-77-220.
\newblock Technical report, Doty Associates, June 1977.

\bibitem{ingold13}
Dan Ingold, Barry Boehm, and Supannika Koolmanojwong.
\newblock A model for estimating agile project process and schedule
  acceleration.
\newblock In {\em ICSSP 2013}, pages 29--35, 2013.

\bibitem{jensen83}
R.~Jensen.
\newblock An improved macrolevel software development resource estimation
  model.
\newblock In {\em 5th ISPA Conference}, pages 88--92, April 1983.

\bibitem{jorg15}
M.~Jorgensen.
\newblock The world is skewed: Ignorance, use, misuse, misunderstandings, and
  how to improve uncertainty analyses in software development projects, 2015.
\newblock CREST workshop, 2015, http://goo.gl/0wFHLZ.

\bibitem{jorgensen09}
M.~J{\o}rgensen and T.M. Gruschke.
\newblock The impact of lessons-learned sessions on effort estimation and
  uncertainty assessments.
\newblock {\em Software Engineering, IEEE Transactions on}, 35(3):368 --383,
  May-June 2009.

\bibitem{jorgensen05}
M.~J{\o}rgensen and M.~Shepperd.
\newblock A systematic review of software development cost estimation studies,
  January 2007.
\newblock Available from
  \url{http://www.simula.no/departments/engineering/publications/J{\o}rgensen.2005.12}.

\bibitem{Jorgensen2004}
Magne Jorgensen.
\newblock {A review of studies on expert estimation of software development
  effort}.
\newblock {\em Journal of Systems and Software}, 70(1-2):37--60, February 2004.

\bibitem{yang13}
M.~Li K.~Mao, Y.~Yang and M.~Harman.
\newblock Pricing crowdsourcing-based software development tasks.
\newblock In {\em ICSE, New Ideas and Emerging Results}, pages 1205--1208, San
  Francisco, CA, USA, 2013.

\bibitem{kadoda00}
G.~Kadoda, M.~Cartwright, L.~Chen, and M.~Shepperd.
\newblock Experiences using casebased reasoning to predict software project
  effort, 2000.

\bibitem{kampenes07}
Vigdis~By Kampenes, Tore Dyb{\aa}, Jo~Erskine Hannay, and Dag I.~K. Sj{\o}berg.
\newblock A systematic review of effect size in software engineering
  experiments.
\newblock {\em Information {\&} Software Technology}, 49(11-12):1073--1086,
  2007.

\bibitem{keung2008a}
Jacky~Wai Keung.
\newblock Empirical evaluation of analogy-x for software cost estimation.
\newblock In {\em ESEM '08: International Symposium on Empirical Software
  Engineering and Measurement}, pages 294--296, New York, NY, USA, 2008. ACM.

\bibitem{keung2008c}
Jacky~Wai Keung and Barbara Kitchenham.
\newblock Experiments with analogy-x for software cost estimation.
\newblock In {\em ASWEC '08: Proceedings of the 19th Australian Conference on
  Software Engineering}, pages 229--238, Washington, DC, USA, 2008. IEEE
  Computer Society.

\bibitem{keung2008b}
Jacky~Wai Keung, Barbara~A. Kitchenham, and David~Ross Jeffery.
\newblock Analogy-x: Providing statistical inference to analogy-based software
  cost estimation.
\newblock {\em IEEE Trans. Softw. Eng.}, 34(4):471--484, 2008.

\bibitem{Kirsopp2002}
C.~Kirsopp and M.~Shepperd.
\newblock {Making inferences with small numbers of training sets}.
\newblock {\em IEEE Proc.}, 149, 2002.

\bibitem{koc11b}
E.~Kocaguneli, T.~Menzies, A.~Bener, and J.~Keung.
\newblock Exploiting the essential assumptions of analogy-based effort
  estimation.
\newblock {\em IEEE Transactions on Software Engineering}, 28:425--438, 2012.
\newblock Available from \url{http://menzies.us/pdf/11teak.pdf}.

\bibitem{koc11a}
E.~Kocaguneli, T.~Menzies, and J.W. Keung.
\newblock On the value of ensemble effort estimation.
\newblock {\em Software Engineering, IEEE Transactions on}, 38(6):1403--1416,
  Nov 2012.

\bibitem{me13a}
Ekrem Kocaguneli, Tim Menzies, Jacky Keung, David Cok, and Ray Madachy.
\newblock Active learning and effort estimation: Finding the essential content
  of software effort estimation data.
\newblock {\em IEEE Transactions on Software Engineering}, 39(8):1040--1053,
  2013.

\bibitem{kocaguneli2014transfer}
Ekrem Kocaguneli, Tim Menzies, and Emilia Mendes.
\newblock Transfer learning in effort estimation.
\newblock {\em Empirical Software Engineering}, pages 1--31, 2014.

\bibitem{kocharm13}
Ekrem Kocaguneli, Thomas Zimmermann, Christian Bird, Nachiappan Nagappan, and
  Tim Menzies.
\newblock Distributed development considered harmful?
\newblock In {\em ICSE}, pages 882--890, 2013.

\bibitem{Li2006}
Jingzhou Li and Guenther Ruhe.
\newblock A comparative study of attribute weighting heuristics for effort
  estimation by analogy.
\newblock {\em International Symposium on Empirical Software Engineering},
  page~74, 2006.

\bibitem{Li2007}
Jingzhou Li and Guenther Ruhe.
\newblock Decision support analysis for software effort estimation by analogy.
\newblock In {\em PROMISE '07: Proceedings of the Third International Workshop
  on Predictor Models in Software Engineering}, page~6, 2007.

\bibitem{Li2008}
Jingzhou Li and Guenther Ruhe.
\newblock Analysis of attribute weighting heuristics for analogy-based software
  effort estimation method aqua+.
\newblock {\em Empirical Softw. Engg.}, 13:63--96, February 2008.

\bibitem{Li2009a}
Y.~Li, M.~Xie, and Goh T.
\newblock A study of the non-linear adjustment for analogy based software cost
  estimation.
\newblock {\em Empirical Software Engineering}, pages 603--643, 2009.

\bibitem{lokan06}
C.~Lokan and E.~Mendes.
\newblock Cross-company and single-company effort models using the isbsg
  database: a further replicated study.
\newblock In {\em The ACM-IEEE International Symposium on Empirical Software
  Engineering, November 21-22, Rio de Janeiro}, 2006.

\bibitem{lokan09}
C.~Lokan and E.~Mendes.
\newblock Applying moving windows to software effort estimation.
\newblock In {\em Empirical Software Engineering and Measurement, 2009. ESEM
  2009. 3rd International Symposium on}, pages 111--122, 2009.

\bibitem{me12d}
Tim Menzies, Andrew Butcher, David~R. Cok, Andrian Marcus, Lucas Layman,
  Forrest Shull, Burak Turhan, and Thomas Zimmermann.
\newblock Local versus global lessons for defect prediction and effort
  estimation.
\newblock {\em IEEE Trans. Software Eng.}, 39(6):822--834, 2013.
\newblock Available from \url{http://menzies.us/pdf/12localb.pdf}.

\bibitem{me06d}
Tim Menzies, Zhihao Chen, Jairus Hihn, and Karen Lum.
\newblock Selecting best practices for effort estimation.
\newblock {\em IEEE Transactions on Software Engineering}, November 2006.
\newblock Available from \url{http://menzies.us/pdf/06coseekmo.pdf}.

\bibitem{me07e}
Tim Menzies, Alex Dekhtyar, Justin Distefano, and Jeremy Greenwald.
\newblock Problems with precision.
\newblock {\em IEEE Transactions on Software Engineering}, September 2007.
\newblock \url{http://menzies.us/pdf/07precision.pdf}.

\bibitem{Minku16b}
Tim Menzies, Ekrem Kocag{\"{u}}neli, Leandro Minku, Fayola Peters, and Burak
  Turhan.
\newblock {Chapter 20 - Ensembles of Learning Machines}.
\newblock In {\em Sharing Data and Models in Software Engineering}, pages
  239--265. 2015.

\bibitem{menzies13err}
Tim Menzies, Fayola Peters, and Andrian Marcus.
\newblock Ooops... (errata report for ``{B}etter {C}ross-{C}ompany
  {L}earning'').
\newblock In {\em MSR'13}, 2013.
\newblock http://www.slideshare.net/timmenzies/msr13-mistake.

\bibitem{me04h}
Tim Menzies, D.~Port, Z.~Chen, J.~Hihn, and S.~Stukes.
\newblock Validation methods for calibrating software effort models.
\newblock In {\em Proceedings, ICSE}, 2005.
\newblock Available from \url{http://menzies.us/pdf/04coconut.pdf}.

\bibitem{me12a}
Tim Menzies and Martin Shepperd.
\newblock Special issue on repeatable results in software engineering
  prediction.
\newblock {\em Empirical Software Engineering}, 17(1-2):1--17, 2012.

\bibitem{miller02}
A.~Miller.
\newblock {\em Subset Selection in Regression (second edition)}.
\newblock Chapman \& Hall, 2002.

\bibitem{Minku2011}
Leandro~L. Minku and Xin Yao.
\newblock {A principled evaluation of ensembles of learning machines for
  software effort estimation}.
\newblock In {\em Industrial Management {\&} Data Systems}, volume 106, pages
  9:1--9:10, 2011.

\bibitem{Minku2013}
Leandro~L. Minku and Xin Yao.
\newblock {Ensembles and locality: Insight on improving software effort
  estimation}.
\newblock In {\em Information and Software Technology}, volume~55, pages
  1512--1528, 2013.

\bibitem{minku14}
Leandro~L. Minku and Xin Yao.
\newblock How to make best use of cross-company data in software effort
  estimation?
\newblock In {\em ICSE'14}, pages 446--456, 2014.

\bibitem{mittas13}
Nikolaos Mittas and Lefteris Angelis.
\newblock Ranking and clustering software cost estimation models through a
  multiple comparisons algorithm.
\newblock {\em IEEE Trans. Software Eng.}, 39(4):537--551, 2013.

\bibitem{molokk08}
Kjetil Molokken-Pstvold, Nils~Christian Haugen, and Hans~Christian Benestad.
\newblock Using planning poker for combining expert estimates in software
  projects.
\newblock {\em Journal of Systems and Software}, 81:2106–2117, December 2008.

\bibitem{Murphy-Hill2012}
Emerson Murphy-Hill, Chris Parnin, and Andrew~P. Black.
\newblock {How We Refactor, and How We Know It}.
\newblock {\em IEEE Transactions on Software Engineering}, 38(1):5--18, 2012.

\bibitem{Myrtveit}
Ingunn Myrtveit, Erik Stensrud, and Martin Shepperd.
\newblock Reliability and validity in comparative studies of software
  prediction models.
\newblock {\em IEEE Trans. Softw. Eng.}, 31(5):380--391, May 2005.

\bibitem{papa13}
Vasil Papakroni.
\newblock Data carving: Identifying and removing irrelevancies in the data.
\newblock Master's thesis, Lane Department of Computer Science and Electrical
  Engineering, West Virginia Unviersity, 2013.

\bibitem{park88}
R.~Park.
\newblock The central equations of the price software cost model.
\newblock In {\em 4th COCOMO Users Group Meeting}, November 1988.

\bibitem{passos11}
Carol Passos, Ana~Paula Braun, Daniela~S. Cruzes, and Manoel Mendonca.
\newblock Analyzing the impact of beliefs in software project practices.
\newblock In {\em ESEM'11}, 2011.

\bibitem{popper63}
K.R. Popper.
\newblock {\em Conjectures and Refutations,}.
\newblock Routledge and Kegan Paul, 1963.

\bibitem{posnet11}
D.~Posnett, V.~Filkov, and P.~Devanbu.
\newblock Ecological inference in empirical software engineering.
\newblock In {\em Proceedings of ASE'11}, 2011.

\bibitem{putnam76}
L.~Putnam.
\newblock A macro-estimating methodology for software development.
\newblock In {\em Proceedings, IEEE COMPCON76 Fall}, pages 38--43, September
  1976.

\bibitem{Scanniello2013}
Giuseppe Scanniello, Carmine Gravino, Andrian Marcus, and Tim Menzies.
\newblock Class level fault prediction using software clustering.
\newblock In {\em Automated Software Engineering (ASE), 2013 IEEE/ACM 28th
  International Conference on}, pages 640--645. IEEE, 2013.

\bibitem{shaw01}
Mary Shaw.
\newblock The coming-of-age of software architecture research.
\newblock In {\em Proceedings of the 23rd International Conference on Software
  Engineering}, ICSE '01, pages 656--, Washington, DC, USA, 2001. IEEE Computer
  Society.

\bibitem{shepperd97}
M.~Shepperd and C.~Schofield.
\newblock Estimating software project effort using analogies.
\newblock {\em IEEE Transactions on Software Engineering}, 23(12), November
  1997.
\newblock Available from \url{http://www.utdallas.edu/~rbanker/SE_XII.pdf}.

\bibitem{shepperd12a}
Martin~J. Shepperd and Steven~G. MacDonell.
\newblock Evaluating prediction systems in software project estimation.
\newblock {\em Information {\&} Software Technology}, 54(8):820--827, 2012.

\bibitem{CLCS03}
Spareref.com.
\newblock Nasa to shut down checkout \& launch control system, August 26, 2002.
\newblock \url{http://www.spaceref.com/news/viewnews.html?id=475}.

\bibitem{stanley2013predicting}
Clayton Stanley and Michael~D Byrne.
\newblock Predicting tags for stackoverflow posts.
\newblock In {\em Proceedings of ICCM}, volume 2013, 2013.

\bibitem{valerdi11}
R.~Valerdi.
\newblock Convergence of expert opinion via the wideband delphi method: An
  application in cost estimation models.
\newblock In {\em Incose International Symposium, Denver, USA}, 2011.
\newblock Available from http://goo.gl/Zo9HT.

\bibitem{Walkerden1999}
Fiona Walkerden and Ross Jeffery.
\newblock An empirical study of analogy-based software effort estimation.
\newblock {\em Empirical Softw. Engg.}, 4(2):135--158, 1999.

\bibitem{watson77}
C.~Walston and C.~Felix.
\newblock A method of programming measurement and estimation.
\newblock {\em IBM Systems Journal}, 16(1):54--77, 1977.

\bibitem{whigham15}
Peter~A. Whigham, Caitlin~A. Owen, and Stephen~G. Macdonell.
\newblock A baseline model for software effort estimation.
\newblock {\em ACM Trans. Softw. Eng. Methodol.}, 24(3):20:1--20:11, May 2015.

\bibitem{wol74}
R.~Wolverton.
\newblock The cost of developing large-scale software.
\newblock {\em IEEE Trans. Computers}, pages 615--636, June 1974.

\end{thebibliography}
\balance

\end{document}